\documentclass[aps,prd,twocolumn,superscriptaddress,showpacs,preprintnumbers,amsmath,amssymb]{revtex4-1}

\usepackage{graphicx}
\usepackage{dcolumn}
\usepackage{bm}
\usepackage[mathlines]{lineno}
\usepackage{subfigure}
\usepackage{multirow}

\begin{document}

\preprint{\vbox{ \hbox{   }
			      \hbox{Belle Preprint 2014-1}
                  \hbox{KEK Preprint 2013-60}
                  \hbox{Intended for {\it PRL}}
}}

\title{Measurement of the Lepton Forward-Backward Asymmetry \\
in $B \rightarrow X_s \ell^+ \ell^-$ Decays
with a Sum of Exclusive Modes
}



\noaffiliation
\affiliation{University of the Basque Country UPV/EHU, 48080 Bilbao}
\affiliation{Beihang University, Beijing 100191}
\affiliation{University of Bonn, 53115 Bonn}
\affiliation{Budker Institute of Nuclear Physics SB RAS and Novosibirsk State University, Novosibirsk 630090}
\affiliation{Faculty of Mathematics and Physics, Charles University, 121 16 Prague}
\affiliation{Chiba University, Chiba 263-8522}
\affiliation{University of Cincinnati, Cincinnati, Ohio 45221}
\affiliation{Deutsches Elektronen--Synchrotron, 22607 Hamburg}
\affiliation{Justus-Liebig-Universit\"at Gie\ss{}en, 35392 Gie\ss{}en}
\affiliation{Hanyang University, Seoul 133-791}
\affiliation{University of Hawaii, Honolulu, Hawaii 96822}
\affiliation{High Energy Accelerator Research Organization (KEK), Tsukuba 305-0801}
\affiliation{IKERBASQUE, Basque Foundation for Science, 48011 Bilbao}
\affiliation{Indian Institute of Technology Guwahati, Assam 781039}
\affiliation{Indian Institute of Technology Madras, Chennai 600036}
\affiliation{Institute of High Energy Physics, Chinese Academy of Sciences, Beijing 100049}
\affiliation{Institute of High Energy Physics, Vienna 1050}
\affiliation{Institute for High Energy Physics, Protvino 142281}
\affiliation{INFN - Sezione di Torino, 10125 Torino}
\affiliation{J. Stefan Institute, 1000 Ljubljana}
\affiliation{Kanagawa University, Yokohama 221-8686}
\affiliation{Institut f\"ur Experimentelle Kernphysik, Karlsruher Institut f\"ur Technologie, 76131 Karlsruhe}
\affiliation{Kavli Institute for the Physics and Mathematics of the Universe (WPI), University of Tokyo, Kashiwa 277-8583}
\affiliation{Korea Institute of Science and Technology Information, Daejeon 305-806}
\affiliation{Korea University, Seoul 136-713}
\affiliation{Kyungpook National University, Daegu 702-701}
\affiliation{\'Ecole Polytechnique F\'ed\'erale de Lausanne (EPFL), Lausanne 1015}
\affiliation{Faculty of Mathematics and Physics, University of Ljubljana, 1000 Ljubljana}
\affiliation{Luther College, Decorah, Iowa 52101}
\affiliation{University of Maribor, 2000 Maribor}
\affiliation{Max-Planck-Institut f\"ur Physik, 80805 M\"unchen}
\affiliation{School of Physics, University of Melbourne, Victoria 3010}
\affiliation{Moscow Physical Engineering Institute, Moscow 115409}
\affiliation{Moscow Institute of Physics and Technology, Moscow Region 141700}
\affiliation{Graduate School of Science, Nagoya University, Nagoya 464-8602}
\affiliation{Kobayashi-Maskawa Institute, Nagoya University, Nagoya 464-8602}
\affiliation{Nara Women's University, Nara 630-8506}
\affiliation{National Central University, Chung-li 32054}
\affiliation{National United University, Miao Li 36003}
\affiliation{Department of Physics, National Taiwan University, Taipei 10617}
\affiliation{H. Niewodniczanski Institute of Nuclear Physics, Krakow 31-342}
\affiliation{Nippon Dental University, Niigata 951-8580}
\affiliation{Niigata University, Niigata 950-2181}
\affiliation{University of Nova Gorica, 5000 Nova Gorica}
\affiliation{Osaka City University, Osaka 558-8585}
\affiliation{Pacific Northwest National Laboratory, Richland, Washington 99352}
\affiliation{Panjab University, Chandigarh 160014}
\affiliation{Peking University, Beijing 100871}
\affiliation{University of Pittsburgh, Pittsburgh, Pennsylvania 15260}
\affiliation{RIKEN BNL Research Center, Upton, New York 11973}
\affiliation{University of Science and Technology of China, Hefei 230026}
\affiliation{Soongsil University, Seoul 156-743}
\affiliation{Sungkyunkwan University, Suwon 440-746}
\affiliation{School of Physics, University of Sydney, NSW 2006}
\affiliation{Department of Physics, Faculty of Science, University of Tabuk, Tabuk 71451}
\affiliation{Tata Institute of Fundamental Research, Mumbai 400005}
\affiliation{Excellence Cluster Universe, Technische Universit\"at M\"unchen, 85748 Garching}
\affiliation{Toho University, Funabashi 274-8510}
\affiliation{Tohoku Gakuin University, Tagajo 985-8537}
\affiliation{Tohoku University, Sendai 980-8578}
\affiliation{Department of Physics, University of Tokyo, Tokyo 113-0033}
\affiliation{Tokyo Institute of Technology, Tokyo 152-8550}
\affiliation{Tokyo Metropolitan University, Tokyo 192-0397}
\affiliation{Tokyo University of Agriculture and Technology, Tokyo 184-8588}
\affiliation{University of Torino, 10124 Torino}
\affiliation{CNP, Virginia Polytechnic Institute and State University, Blacksburg, Virginia 24061}
\affiliation{Wayne State University, Detroit, Michigan 48202}
\affiliation{Yamagata University, Yamagata 990-8560}
\affiliation{Yonsei University, Seoul 120-749}
  \author{Y.~Sato}\affiliation{Tohoku University, Sendai 980-8578}\affiliation{Kobayashi-Maskawa Institute, Nagoya University, Nagoya 464-8602} 
  \author{A.~Ishikawa}\affiliation{Tohoku University, Sendai 980-8578} 
  \author{H.~Yamamoto}\affiliation{Tohoku University, Sendai 980-8578} 
  \author{A.~Abdesselam}\affiliation{Department of Physics, Faculty of Science, University of Tabuk, Tabuk 71451} 
  \author{I.~Adachi}\affiliation{High Energy Accelerator Research Organization (KEK), Tsukuba 305-0801} 
  \author{K.~Adamczyk}\affiliation{H. Niewodniczanski Institute of Nuclear Physics, Krakow 31-342} 
  \author{H.~Aihara}\affiliation{Department of Physics, University of Tokyo, Tokyo 113-0033} 
  \author{D.~M.~Asner}\affiliation{Pacific Northwest National Laboratory, Richland, Washington 99352} 
  \author{V.~Aulchenko}\affiliation{Budker Institute of Nuclear Physics SB RAS and Novosibirsk State University, Novosibirsk 630090} 
  \author{T.~Aushev}\affiliation{Moscow Institute of Physics and Technology, Moscow Region 141700} 
% ...
\collaboration{The Belle Collaboration}

\date{\today}

\begin{abstract}
We report the first measurement of
the lepton forward-backward asymmetry ${\cal A}_{\rm FB}$
as a function of the squared four-momentum of the dilepton system, $q^2$,
for the electroweak penguin process $B \rightarrow X_s \ell^+ \ell^-$
with a sum of exclusive final states,
where $\ell$ is an electron or a muon
and $X_s$ is a hadronic recoil system with an  $s$ quark.
The results are based on a data sample
containing $772\times10^6$ $B\bar{B}$ pairs 
recorded at the $\Upsilon(4S)$ resonance
with the Belle detector at the KEKB $e^+ e^-$ collider.
${\cal A}_{\rm FB}$ for the inclusive $B \rightarrow X_s \ell^+ \ell^-$
is extrapolated from the sum of 10 exclusive $X_s$ states
whose invariant mass is less than 2 GeV/$c^2$.
For $q^2 > 10.2$ GeV$^2$/$c^2$,
${\cal A}_{\rm FB} < 0$ is excluded at the 2.3$\sigma$ level,
where $\sigma$ is the standard deviation.
For $q^2 < 4.3$ GeV$^2$/$c^2$,
the result is within 1.8$\sigma$
of the Standard Model theoretical expectation.
\end{abstract}


\maketitle

\section{INTRODUCTION}
In the Standard Model (SM),
quark-level flavor-changing neutral current 
$b \rightarrow s \ell^+ \ell^-$ 
decays \cite{CHARGE_CONJUGATION} are allowed
at higher order via the electroweak loop (penguin)
and $W^+ W^-$ box diagrams.
The corresponding decay amplitude
can be expressed
via the Operator Product Expansion \cite{OPE}
in terms of the effective Wilson coefficients
for the electromagnetic penguin, $C_7^{\rm eff}$,
and the vector and axial-vector electroweak contributions,
$C_9^{\rm eff}$ and $C_{10}^{\rm eff}$, respectively
\cite{WC_XSLL}.
If physics beyond the SM contributes to $b \rightarrow s \ell^+ \ell^-$ decays,
then the effective Wilson coefficients are expected to differ 
from the SM expectations.
Therefore the decay rate and angular distributions
of $b \rightarrow s \ell^+ \ell^-$ decays
constitute good probes to search for new physics
\cite{THEORY_XSLL}.

Inclusive measurements of the $b \rightarrow s \ell^+ \ell^-$ process are preferable
to exclusive measurements
because of lower theoretical uncertainties,
although they are experimentally more challenging.
The branching fraction for inclusive $B \rightarrow X_s \ell^+ \ell^-$,
where $B$ is either $\bar{B}^0$ or $B^-$,
$\ell$ is either an electron or a muon,
and $X_s$ is a hadronic recoil system with an $s$ quark, has been measured
by Belle \cite{BR_XSLL_BELLE}
and
\mbox{\sl B\hspace{-0.4em} {\small\sl A}\hspace{-0.37em} \sl B\hspace{-0.4em}
{\small\sl A\hspace{-0.02em}R}}
\cite{BR_XSLL_BABAR}.
Both results are consistent
with the SM prediction.
The lepton forward-backward asymmetry,
defined as
\begin{eqnarray}
  &&{\cal A}_{\rm FB}(q^2_{\rm min},q^2_{\rm max}) \nonumber \\
  && \hspace{2.0em} =
   { \int_{q^2_{\rm min}}^{q^2_{\rm max}} dq^2 \int_{-1}^1 d\cos\theta
   \;{\rm sgn}(\cos\theta) {d^2 \Gamma \over dq^2 d\cos\theta} \over
     \int_{q^2_{\rm min}}^{q^2_{\rm max}} dq^2 \int_{-1}^1 d\cos\theta
       {d^2 \Gamma \over dq^2 d\cos\theta} }
\label{eq:afb}
\end{eqnarray}
is considered to have different
and greater sensitivity to physics beyond the SM
than the branching fraction \cite{IMDEP_THEORY,IMDEP_THEORY_NEW}.
Here,
$q^2$ is the squared four-momentum of the dilepton system
and
$\theta$ is the angle between
the $\ell^+ (\ell^-)$ and the $B$ meson momentum
in the $\ell^+ \ell^-$ center-of-mass frame
in $\bar{B}^0$ or $B^-$ ($B^0$ or $B^+$) decays.
Although ${\cal A}_{\rm FB}$ in exclusive $B \rightarrow K^{(*)} \ell^+ \ell^-$ has been measured
by Belle \cite{AFB_XSLL_BELLE},
\mbox{\sl B\hspace{-0.4em} {\small\sl A}\hspace{-0.37em} \sl B\hspace{-0.4em}
{\small\sl A\hspace{-0.02em}R}}
\cite{AFB_XSLL_BABAR},
CDF \cite{AFB_XSLL_CDF}, LHCb \cite{AFB_XSLL_LHCb}
and CMS \cite{AFB_XSLL_CMS},
${\cal A}_{\rm FB}$ in inclusive $B \rightarrow X_s \ell^+ \ell^-$
is yet to be measured.
At lowest order,  the numerator in Eq.~\ref{eq:afb}
for inclusive $B \rightarrow X_s \ell^+ \ell^-$
can be written~\cite{AFB_XSLL_EQ} as a function of $q^2$
\begin{eqnarray}
&\int^{1}_{-1}&{\rm sgn}(\cos\theta) \frac{d^2 \Gamma}{dq^2d\cos\theta}d\cos\theta \nonumber \\
&=&
-3 \Gamma_0 m_b^3 c^8
(1-s)^2 s C_{10} {\rm Re}
\left( 
C_9 + \frac{2}{s} C_7
\right),
\end{eqnarray}
where
$m_b$ is the $b$-quark mass,
$s = q^2/(m_b^2c^2)$, and
$\Gamma_0 = \frac{G_F^2}{48\hbar^6 c^6 \pi^3} \frac{\alpha_{\rm em}}{16\pi^2} | V_{tb} V_{ts}^*|^2$.
Here,
$G_F$ is the Fermi coupling constant,
$V_{tb}$ and $V_{ts}$ are Cabibbo-Kobayashi-Maskawa matrix elements~\cite{CKM},
and $\alpha_{\rm em}$ is the fine-structure constant.

We report the first measurement
of the lepton forward-backward asymmetry
for inclusive $B \rightarrow X_s \ell^+ \ell^-$,
which is extrapolated from
the sum of 10 exclusive $X_s$ states
with an invariant mass $M_{X_s} < 2.0$  GeV$/c^2$,
corresponding to 50\% of the inclusive rate.
We also report this asymmetry
for the subsamples of
$B \rightarrow K^{(*)} \ell^+ \ell^-$ with the $X_s$ invariant mass $M_{X_s} < 1.1$ GeV/$c^2$
and
$B \rightarrow X_s \ell^+ \ell^-$ with $M_{X_s} > 1.1$ GeV/$c^2$,
where this asymmetry for $B \rightarrow K \ell^+ \ell^-$ is expected to be zero in the SM.
We assume that ${\cal A}_{\rm FB}$ is independent of lepton flavor.
When the final state $X_s$ is not a $K^{(*)}$,
we also assume ${\cal A}_{\rm FB}$ depends neither on $X_s$ nor on the $X_s$ mass.
The results are based on the full $\Upsilon(4S)$ data sample
containing $772 \times 10^6$ $B \bar{B}$ pairs
recorded with the Belle detector \cite{BELLE}
at the KEKB $e^+ e^-$ collider \cite{KEKB}.

\section{DETECTOR}
The Belle detector is a general-purpose magnetic spectrometer
which consists of
a silicon vertex detector (SVD),
a 50-layer central drift chamber (CDC),
an array of aerogel threshold Cherenkov counters (ACC),
time-of-flight scintillation counters (TOF),
and an electromagnetic calorimeter (ECL) comprised of CsI(Tl) crystals.
The devices are located inside a superconducting solenoid coil
that provides a 1.5 T magnetic field.
An iron flux-return located outside the coil
is instrumented to detect $K_L^0$ mesons
and to identify muons (KLM).
The detector is described in detail elsewhere \cite{BELLE}.

\section{SIGNAL MODEL}
We study the acceptance for
$B \rightarrow X_s \ell^+ \ell^-$ via Monte Carlo (MC) simulation.
For this simulation,
we use a sum of 
exclusive $B \rightarrow K^{(*)} \ell^+ \ell^-$ events and
non-resonant $B \rightarrow X_s \ell^+ \ell^-$ events
with $M_{X_s} >$  1.1~GeV/$c^2$.
The former are generated according to
Refs.~\cite{THEORY_XSLL,EVTGEN_KLL},
while the latter are generated using a model based on
Refs.~\cite{THEORY_XSLL,EVTGEN_XSLL}
and the Fermi motion model
of Ref.~\cite{FERMIMOTION}.
The two MC samples are mixed
assuming the measured branching fractions \cite{HFAG}.

\section{EVENT SELECTION}
Charged tracks are reconstructed with
the SVD and CDC,
and the tracks other than $K_S^0 \rightarrow \pi^+ \pi^-$ daughters are required to originate from
the interaction region.
Electrons are identified
by a combination of
the specific ionization ($dE/dx$) in the CDC,
the ratio of the cluster energy in the ECL
to the track momentum measured with the SVD and CDC,
the response of the ACC,
the shower shape in the ECL,
and position matching between the shower and the track.
Muons are identified
by the track penetration depth and hit scatter in the KLM.
Electrons and muons are required to have momenta
greater than 0.4 GeV/$c$ and 0.8 GeV/$c$, respectively.
To recover bremsstrahlung photons from leptons,
we add the four-momentum of each photon detected
within 0.05 rad of the original track direction.
Charged kaons are identified
by combining information from
the $dE/dx$ in the CDC, 
the flight time measured with the TOF,
and the response of the ACC \cite{PID}.
We select electron, muon, and kaon candidate tracks in turn,
while the remaining tracks are assumed to be charged pions.

$K_S^0$ candidates are formed
by combining two oppositely charged tracks,
assuming both are pions
with requirements on their invariant mass, flight length,
and consistency between the $K_S^0$ momentum direction and vertex position.
Neutral pion candidates are formed
from pairs of photons
that have an invariant mass within 10 MeV/$c^2$
of the nominal $\pi^0$ mass,
where photons are measured as an energy cluster
in the ECL with no associated charged tracks.
Neutral pions and their photon daughters are required
to have an energy greater than
400 MeV and 50 MeV, respectively.
A mass-constrained fit is then performed to obtain the $\pi^0$ momentum.

\begin{table}[htb]
\caption{
The 18 hadronic final states used to reconstruct $X_s$.
The 8 final states enclosed in parentheses are not used for the measurement of ${\cal A}_{\rm FB}$.
}
\label{tab:rec_mode}
\scalebox{0.91}{
\begin{tabular}{@{\hspace{-0.06em}}l@{\hspace{-0.06em}}l@{\hspace{-0.06em}}|@{\hspace{0.00em}}l@{\hspace{-0.10em}}l@{\hspace{-0.05em}}} \hline \hline
\multicolumn{2}{c|}{$\bar{B}^0$ decays} & \multicolumn{2}{c}{$B^-$ decays}\\ \hline
                            & ($K_S^0$)                       & \ $K^-$                    &                                 \\
 $K^-\pi^+$                 & ($K_{S}^0\pi^{0}$)              & \ $K^-\pi^0$               & \ $K_{S}^0\pi^-$                \\
 $K^-\pi^+\pi^0$            & ($K_{S}^0\pi^-\pi^+$)           & \ $K^-\pi^+\pi^-$          & \ $K_{S}^0\pi^-\pi^{0}$         \\
 $K^-\pi^+\pi^-\pi^+$       & ($K_{S}^0\pi^-\pi^+\pi^{0}$)    & \ $K^-\pi^+\pi^-\pi^0$     & \ $K_{S}^0\pi^-\pi^+\pi^-$      \\
($K^-\pi^+\pi^-\pi^+\pi^0$) & ($K_{S}^0\pi^-\pi^+\pi^-\pi^+$) & ($K^-\pi^+\pi^-\pi^+\pi^-$)& ($K_{S}^0\pi^-\pi^+\pi^-\pi^0$) \\ \hline \hline
\end{tabular}}
\end{table}

We reconstruct $X_s$ from 18 hadronic final states
(see Table~\ref{tab:rec_mode}),
that consist of one $K^{\pm}$ or $K_S^0$
and up to four pions,
of which at most one can be neutral.
To reject a large part of the combinatorial background,
we require $M_{X_s} < 2$ GeV/$c^2$,
which preserves 91\% of signal.

We combine the $X_s$ with two oppositely charged leptons
to form a $B$ meson candidate.
To identify the signal,
we use two kinematic variables defined
in the $\Upsilon(4S)$ rest frame:
the beam-energy constrained mass
$M_{\rm bc} = \sqrt{E_{\rm beam}^{*2} - |\vec{p}_B|^2}$,
and the energy difference
$\Delta E = E_B - E_{\rm beam}^*$,
where $E_{\rm beam}^*$ is the beam energy
and $(\vec{p}_B, E_B)$ is the reconstructed
momentum and energy of the $B$ candidate.
We require $M_{\rm bc} > 5.22$ GeV/$c^2$
and $-100$ MeV $< \Delta E < 50$ MeV
$(-50$ MeV $< \Delta E < 50$ MeV)
for the electron (muon) channel.

To reject large contamination from charmonium backgrounds
$B \rightarrow J/\psi (\psi(2S)) X_s$
followed by
$J/\psi (\psi(2S)) \rightarrow \ell^+ \ell^-$,
we reject events having dilepton invariant mass in the following veto regions:
$-400$ to $150$ MeV/$c^2$ ($-250$ to $100$ MeV/$c^2$)
around the $J/\psi$ mass
and 
$-250$ to $100$ MeV/$c^2$ ($-150$ to $100$ MeV/$c^2$)
around the $\psi(2S)$ mass
for the electron (muon) channel.
In the electron channel,
there is non-negligible peaking background from events
in which
the bremsstrahlung photon recovery fails
and instead the radiated photon together
with another random photon forms a misreconstructed $\pi^0$ as $X_s$'s daughter.
To veto such events,
the $\pi^0$'s photon daughter with the highest energy 
is added in the calculation of the dilepton invariant mass,
and events with invariant mass from $150$ MeV/$c^2$ below to $50$ MeV/$c^2$ above the nominal $J/\psi$ mass
are rejected for the modes involving $\pi^0$.
We also require the dilepton mass to be greater than
0.2 GeV/$c^2$
to remove
the photon conversion and $\pi^0$ Dalitz decays.

\section{BACKGROUND SUPPRESSION}
The main background comes from
random combinations of two semileptonic $B$ or $D$ decays,
which have both large missing energy due to neutrinos,
and displaced origin of leptons from $B$ or $D$ mesons.
The displacement between the two leptons is measured
by the distance $\Delta z_{\ell^+\ell^-}$
between the points of closest approach to the beam axis along the beam direction.
We also use the confidence level of the $B$ vertex ($\mathcal{C}_{\rm{vtx}}$),
constructed from all charged daughter particles except for
$K_S^0$ daughters.
We set requirements on $\Delta z_{\ell^+\ell^-}$ and $\mathcal{C}_{\rm{vtx}}$
to preserve about 79\% of the signal
while rejecting 66\% of the background.
Other background originates from
$e^+ e^- \rightarrow q \bar{q}$
$(q=u,d,s,c)$ continuum events, which can be efficiently suppressed using event shape variables.

To suppress the continuum background and further reduce the semileptonic background,
we employ a neural network based on the software package ``NeuroBayes" \cite{NEUROBAYES}.
The inputs to the network are
(i) a likelihood ratio based on $\Delta E$,
(ii) the cosine of the angle between the $B$ candidate
and the beam axis in the $\Upsilon(4S)$ rest frame,
(iii) $\Delta z_{\ell^+ \ell^-}$,
(iv) $\mathcal{C}_{\rm{vtx}}$,
(v) the total visible energy,
(vi) the missing mass \cite{EVISMMISS},
and
(vii) 17 event shape variables
based on modified Fox-Wolfram moments \cite{Fox}.
For the different types of backgrounds
(semileptonic and continuum),
the neural network is trained separately
and requirements on
two output values are chosen
to maximize the statistical significance.
This optimization is performed separately
for electron and muon channels
and for the regions $M_{X_s} < 1.1$ GeV/$c^2$
and $M_{X_s} > 1.1$ GeV/$c^2$,
and the obtained selection preserves
51\% (63\%) of the signal
while rejecting 98\% (96\%) of the background
for electron (muon) channels. 
According to the MC simulation,
83\% of the remaining background consists of semileptonic events.

The probability of multiple $B$ candidates in a signal event is 8\%
with the average number of $B$ candidates per signal event being 1.1.
When multiple $B$ candidates are found in an event,
we select the most signal-like $B$ candidate
based on the neural network output. 
For the measurement of ${\cal A}_{\rm FB}$,
information on the flavor of the $B$ candidate is necessary.
For $\bar{B}^0$ mesons,
only the self-tagging modes with a $K^-$
are kept,
after selecting one $B$ candidate per event.
We also remove candidates with $X_s$
reconstructed from one kaon plus four pions
because expected signal yields are less than one event. 
Therefore,
we use 10 final states as listed in Table~\ref{tab:rec_mode}
for the $X_s$
to measure ${\cal A}_{\rm FB}$.

\section{MAXIMUM LIKELIHOOD FIT}
To examine the $q^2$ dependence of ${\cal A}_{\rm FB}$,
we divide the data into 4 bins of measured $q^2$:
[0.2, 4.3],
[4.3, 7.3(8.1)],
[10.5(10.2), 11.8(12.5)],
[14.3, 25.0] GeV$^2$/$c^2$
for the electron (muon) channel,
where the gap regions correspond to the veto regions
for charmonium background events.
The bins are numbered in the order of increasing $q^2$;
the lowest $q^2$ for bin number 1,
and the highest for bin number 4.
In order to extract ${\cal A}_{\rm FB}$,
an extended unbinned maximum likelihood fit to four $M_{\rm bc}$ distributions
(positive/negative $\cos\theta$ for electron/muon channel)
is simultaneously performed for each $q^2$ bin.
We also measure ${\cal A}_{\rm FB}$ in the low-$q^2$ region, $1 < q^2 < 6$ GeV$^2/c^2$,
where it is theoretically clean. 

\begin{figure}[htb]
\centering

\subfigure[Electron channel.]{
\includegraphics*[width=7cm]{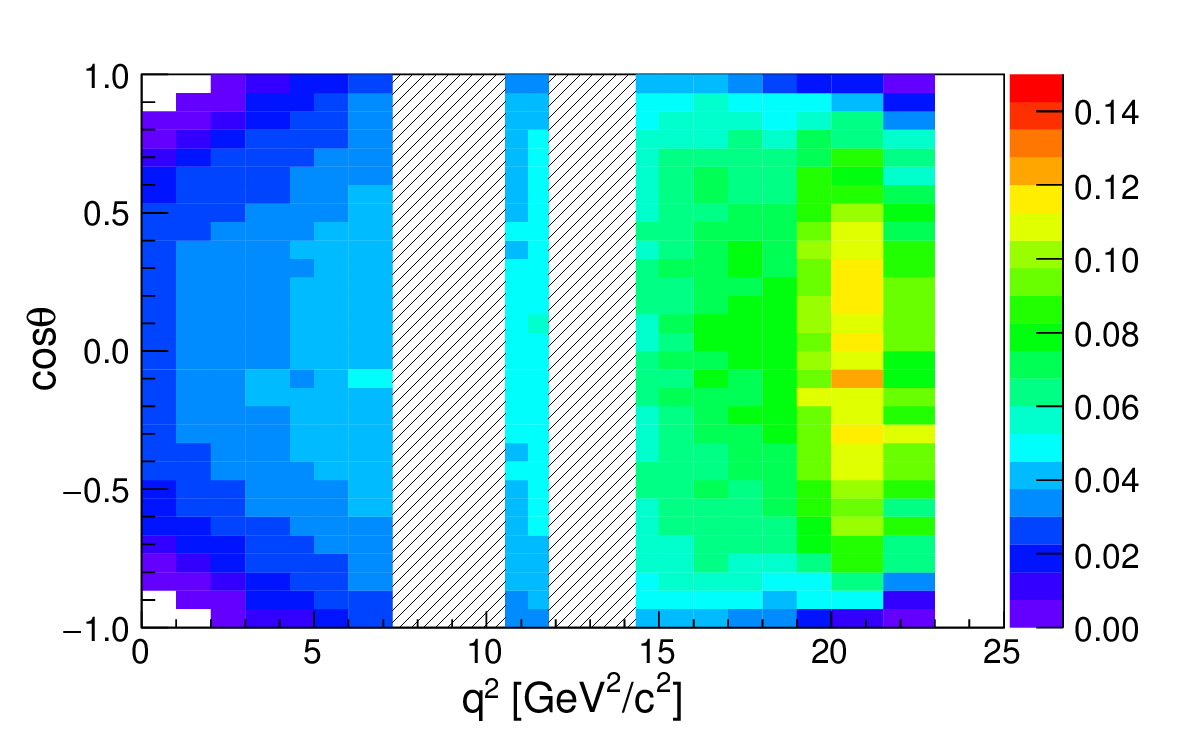}
\label{fig:rec_eff_ee}}
\subfigure[Muon channel.]{
\includegraphics*[width=7cm]{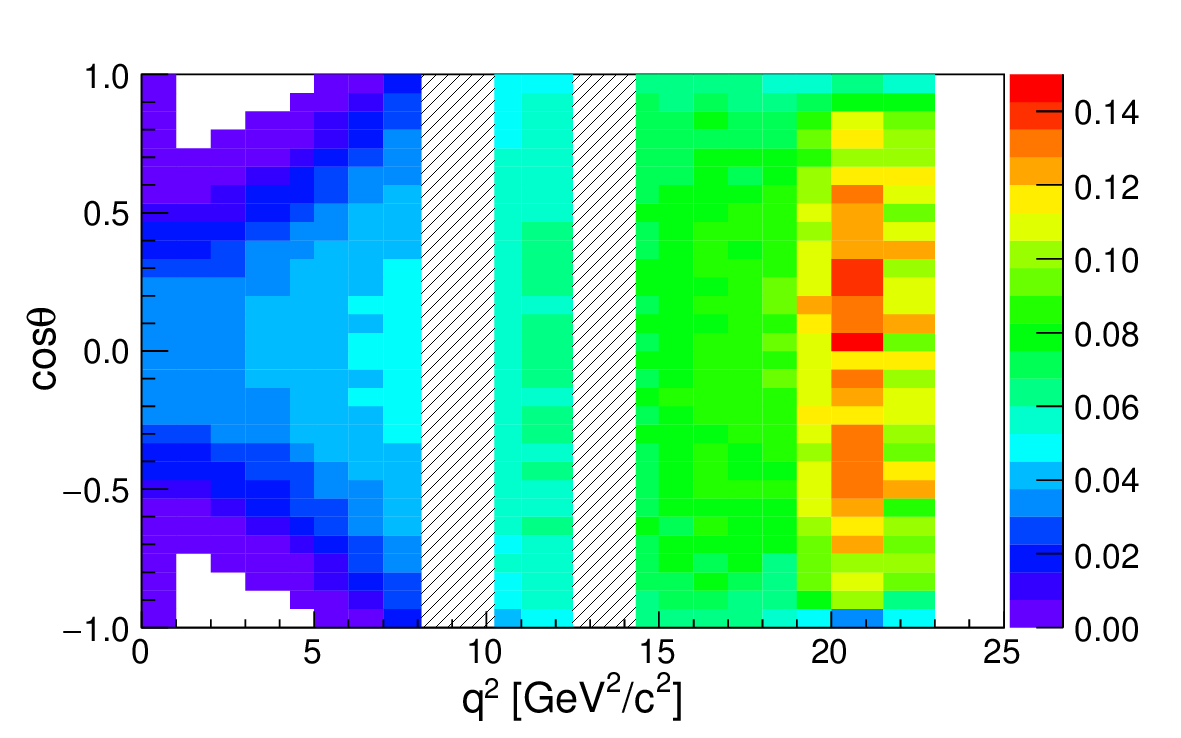}
\label{fig:rec_eff_mm}}
\caption{
Reconstruction efficiency on a plane of $q^2$ and $\cos \theta$ for (a) the electron and (b) the muon channels.
The $J/\psi$ and $\psi(2S)$ veto regions are shown as hatched regions.
\label{fig:rec_eff}
}
\end{figure}

The raw asymmetry ${\cal A}_{\rm FB}^{\rm raw}= \frac{N(\cos\theta>0) - N(\cos\theta<0)}{N(\cos\theta>0) + N(\cos\theta<0)}$,
where $N$ is the observed signal yields,
differs from ${\cal A}_{\rm FB}$
due to the dependence of the signal reconstruction efficiency on $q^2$ and $\cos \theta$.
Figure~\ref{fig:rec_eff} show the reconstruction efficiencies on a plane of $q^2$ and $\cos \theta$.
This pronounced dependence arises from
events with low $q^2$ and high $\cos\theta$
having lepton momenta below the event selection requirements.
We define $\alpha$ as a scaling factor
that relates ${\cal A}_{\rm FB}^{\rm raw}$ to ${\cal A}_{\rm FB}$.
We assume that ${\cal A}_{\rm FB}$ does not depend on the lepton flavor.
However,
${\cal A}_{\rm FB}$ in the second and third $q^2$ bins do differ between electron and muon channels
due to the distinct charmonium-veto regions.
We identify ${\cal A}_{\rm FB}$ as the fit parameter for the $q^2$ regions of the muon channel
and then introduce the scaling factor $\beta$ between the values in the electron and muon channels.
With these factors,
the fit parameter ${\cal A}_{\rm FB}$ is
\begin{eqnarray}
{\cal A}_{\rm FB} &\equiv& {\cal A}_{\rm FB}^{\mu\mu} \nonumber \\
&=& \beta \cdot {\cal A}_{\rm FB}^{ee}, \hbox{ where} \nonumber \\
{\cal A}_{\rm FB}^{\ell \ell} &=& \alpha^{\ell \ell} \cdot {\cal A}_{\rm FB}^{{\rm raw}, \ell \ell}
\hspace{1.0em} (\ell = e, \mu). 
\label{eq:cf}
\end{eqnarray}

To derive $\alpha^{\ell\ell}$ $(\ell = e, \mu)$,
we generate several sets of signal MC samples with various Wilson coefficients ($C_{7}$, $C_{9}$, $C_{10}$),
and calculate ${\cal A}_{\rm FB}^{\ell \ell}$ for each set.
We evaluate ${\cal A}_{\rm FB}^{{\rm raw},\ell \ell}$
using the reconstruction efficiency
as a function of $q^2$ and $\cos \theta$.
We derive $\alpha^{\ell\ell}$
by fitting the relation between ${\cal A}_{\rm FB}^{\ell\ell}$ and
${\cal A}_{\rm FB}^{{\rm raw},\ell \ell}$ to a straight line.
In the first $q^2$ bin,
the quite distinct values of $\alpha$ between electron and muon channels
reflect the different lepton momentum selection criteria. 
To derive $\beta$,
we fit the relation between ${\cal A}_{\rm FB}^{ee}$ and ${\cal A}_{\rm FB}^{\mu \mu}$
in the same way.
The values of $\alpha$ and $\beta$ are summarized in Table~\ref{tab:results}.

\begin{table*}[htb]
\caption{\label{tab:results}Fit results for the five $q^2$ bins.
For ${\cal A}_{\rm FB}$,
the first uncertainty is statistical
and the second uncertainty is systematic.
${\cal A}_{\rm FB}$ values predicted by the SM
\cite{THEORY_XSLL, IMDEP_THEORY}
are also shown with systematic uncertainties.
For the signal yields,
only statistical uncertainties are shown.
The uncertainties of $\alpha$ and $\beta$ are
due to the statistical uncertainties of the MC.
}
\begin{ruledtabular}
\begin{tabular}{c|ccccc}
                                             & 1st $q^2$ bin                    & 2nd $q^2$ bin                       & 3rd $q^2$ bin                         & 4th $q^2$ bin                       & \\ \hline
\multirow{2}{*}{$q^2$ range [GeV$^2$/$c^2$]} & \multirow{2}{*}{[0.2,4.3]} & [4.3,7.3]$_{X_s e^+ e^-}$     & [10.5,11.8]$_{X_s e^+ e^-}$     & \multirow{2}{*}{[14.3, 25.0]} & \multirow{2}{*}{[1.0, 6.0]}\\
                                             &                            & [4.3,8.1]$_{X_s \mu^+ \mu^-}$ & [10.2,12.5]$_{X_s \mu^+ \mu^-}$ &                               & \\ \hline
${\cal A}_{\rm FB}$                          & $0.34  \pm 0.24 \pm 0.03$  & $0.04 \pm 0.31 \pm 0.05$      & $0.28 \pm 0.21 \pm 0.02$        & $0.28 \pm 0.15 \pm 0.02$      & $0.30 \pm 0.24 \pm 0.04$ \\
${\cal A}_{\rm FB}$ (theory)                 & $-0.11 \pm 0.03$           & $0.13 \pm 0.03$               & $0.32 \pm 0.04$                 & $0.40 \pm 0.04$               & $-0.07 \pm 0.04$ \\
$N_{\rm sig}^{ee}$                           & $45.6  \pm 10.9$           & $30.0 \pm  9.2$               & $25.0 \pm  7.0$                 & $39.2 \pm  9.6$               & $50.3 \pm 11.4$ \\
$N_{\rm sig}^{\mu\mu}$                       & $43.4  \pm  9.2$           & $23.9 \pm 10.4$               & $30.7 \pm  9.9$                 & $62.8 \pm 10.4$               & $35.3 \pm  9.2$ \\
$\alpha^{ee}$                                & $1.289 \pm 0.004$          & $1.139 \pm 0.003$             & $1.063 \pm 0.003$               & $1.121 \pm 0.003$             & $1.255 \pm 0.003$ \\
$\alpha^{\mu\mu}$                            & $2.082 \pm 0.010$          & $1.375 \pm 0.003$             & $1.033 \pm 0.003$               & $1.082 \pm 0.003$             & $1.863 \pm 0.006$ \\ 
$\beta$                                      & $1.000$                    & $1.019 \pm 0.003$             & $1.003 \pm 0.000$               & $1.000$                       & $1.000$ \\    
\end{tabular}
\end{ruledtabular}
\end{table*}

The likelihood function
consists of four components:
signal, self cross-feed, combinatorial background, and peaking background.
The signal is modeled with a Gaussian function
with parameters obtained from the $B \rightarrow J/\psi X_s$ data.
The self cross-feed is described by a MC histogram,
where the yield ratio to the signal is fixed
according to the MC expectation.
The combinatorial background is modeled
by an ARGUS function \cite{ARGUS},
where the endpoint is fixed
to the nominal beam energy in the $\Upsilon(4S)$ rest frame,
$E_{\rm beam}^* = 5.289$ GeV.
We have three peaking background sources.
First is charmonium peaking background, $B \rightarrow J/\psi (\psi(2S)) X_s$ decays with
The yields and shape of these charmonium peaking backgrounds
is modeled by histogram shape of charmonium MC samples.
The yields of charmonium peaking background are
estimated to be $0.9 \pm 0.2$ and $2.1 \pm 0.2$ events
in the electron and muon channels, respectively. 
We treat contributions from charmonium resonances higher than $\psi(2S)$ as signal.
Second is $B \rightarrow D^{(*)} n \pi$ $(n > 0)$ decay
with misidentification of two charged pions as two leptons.
The yields and shape of this peaking background
is determined directly from the data
by performing the analysis
without the lepton identification requirements.
Taking the $\pi \rightarrow \ell$ misidentification rates into account,
We estimate this peaking background to be
$0.07 \pm 0.01$ and $5.0 \pm 0.2$ events
in the electron and muon channels, respectively.
Third is $B \rightarrow J/\psi (\psi(2S)) X_s$
with swapped misidentification between a lepton and a pion.
The yields and shape of this peaking background
is determined directly from the data
by performing the analysis
selecting dilepton invariant mass around $J/\psi$ and $\psi(2S)$.
Taking the $\pi \rightarrow \ell$ misidentification rates
and particle identification efficiencies into account,
we estimate this peaking background to be
$0.06 \pm 0.02$ and $4.3 \pm 0.2$ events
in the electron and muon channels, respectively.

\section{SYSTEMATIC UNCERTAINTIES}
To estimate systematic uncertainties,
we repeat the ${\cal A}_{\rm FB}$ fit with varied input parameters
and the resulting change in ${\cal A}_{\rm FB}$
is taken as the systematic uncertainty
for the varied parameter.
Systematic uncertainties for ${\cal A}_{\rm FB}$
are summarized in Table~\ref{tab:sys}.
In the 1st $q^2$ bin,
the dominant systematic uncertainty arises from the translation
of ${\cal A}_{\rm FB}^{\rm raw}$ to ${\cal A}_{\rm FB}$ with $\alpha$ and $\beta$.
Even if a MC sample with a different set
of Wilson coefficients produces
the same values of ${\cal A}_{\rm FB}$,
the ${\cal A}_{\rm FB}^{\rm raw}$ values
and hence the $\alpha$ coefficient may differ.
It gives rise to an uncertainty of the offset in the linear fit.
To estimate this uncertainty,
the relation between ${\cal A}_{\rm FB}^{\rm raw}$ and ${\cal A}_{\rm FB}$
are projected onto the axis perpendicular to the fitted linear line
and fitted by a Gaussian function.
To estimate systematic uncertainties
from the peaking background,
the yield of each such background is varied by its uncertainty.
For the charmonium peaking background,
the yield is varied by $\pm100$\%, conservatively,
because it is determined from MC events.
A possible peaking background from
$B \rightarrow K n\pi \ell \nu$ $(n>0)$,
where one pion is misidentified as a lepton
and the missing neutrino is compensated
by a pion of the other $B$ decay,
is examined.
The number of events
in the whole $q^2$ region is estimated from MC
to be $0.2 \pm 0.6$ $(1.1 \pm 0.7)$ for
electron (muon) channel,
and the resulting systematic error is
${\cal O}(0.001)$.
In the 2nd $q^2$ bin, the systematic uncertainty from charmonium peaking background is dominant.
To estimate the systematic uncertainties from signal modeling,
the related parameters are varied. 
The fraction of $B \rightarrow K^{(*)} \ell^+ \ell^-$
and non-resonant $B \rightarrow X_s \ell^+ \ell^-$
are varied within experimental uncertainties.
$B \rightarrow K^{(*)} \ell^+ \ell^-$ MC samples are generated with different form factors \cite{SYS_FF1, SYS_FF2}.
The Fermi motion parameter is varied
in accordance with
measurements of hadronic moments
in semileptonic $B$ decays \cite{SYS_FERMI_SEMI}
and the photon spectrum
in inclusive $B \rightarrow X_s \gamma$ decays \cite{SYS_FERMI_XSGAMMA}.
The $b$-quark pole mass is varied
by $\pm0.15$ GeV/$c^2$
around 4.80 GeV/$c^2$.
The threshold point of non-resonant $B \rightarrow X_s \ell^+ \ell^-$ events
is varied by $\pm100$ MeV/$c^2$
around $M_{X_s} = 1.1$ GeV/$c^2$.
In the region $M_{X_s} < 1.1$ GeV/$c^2$,
there is possible contamination from the non-resonant $S$-wave component of the $K \pi$ system.
Nevertheless, we find negligible systematic uncertainty from this effect
by adding 5\% contributions of $S$-wave components to the dominant $K^*$
in this $M_{X_s}$ region \cite{SWAVE}.
We check the hadronization process in the non-resonant
$B \rightarrow X_s \ell^+ \ell^-$ events by
comparing the $B \rightarrow J/\psi X_s$ events
in data and MC simulations.
To estimate the systematic uncertainties related to $X_s$ spin components,
we generate non-resonant $B \rightarrow X_s \ell^+ \ell^-$ MC samples with spin 0 and 1
using the form factor for $B \rightarrow K^{(*)} \ell^+ \ell^-$.
In these MC samples, $X_s$ always decays to the two-body $K\pi$ final states to enhance the effect of the $X_s$ spin.
We replace the nominal non-resonant $B \rightarrow X_s \ell^+ \ell^-$ MC samples with these MC samples,
and estimate the systematic uncertainty from the difference between MC samples with spin 0 and 1.
The signal shape parameters are fixed
using the $J/\psi X_s$ data.
The mean and width of the signal Gaussian function are varied
within their uncertainties.
The histogram shape of the self cross-feed background is estimated
from signal MC events.
The entries in the bins are varied
according to a Gaussian distribution
whose standard deviation is the statistical uncertainty of the MC sample.
The total systematic uncertainty is estimated
by summing the above uncertainties in quadrature.

\begin{table*}[htb]
\caption{Summary of systematic uncertainties in the five $q^2$ bins.}
\label{tab:sys}
\begin{tabular}
{c|ccccc}
\hline \hline
Sources of uncertainties                                              & \hspace{1.0em}1st $q^2$ bin\hspace{1.0em} & \hspace{1.0em}2nd $q^2$ bin\hspace{1.0em} & \hspace{1.0em}3rd $q^2$ bin\hspace{1.0em} & \hspace{1.0em}4th $q^2$ bin\hspace{1.0em} &  $1 < q^2 < 6$ GeV$^2$/$c^2$ \\ \hline
Translation from ${\cal A}_{\rm FB}^{\rm raw}$ to ${\cal A}_{\rm FB}$ & 0.019         & 0.013         & 0.007         & 0.003 & 0.020 \\
Peaking background                                                    & 0.004         & 0.050         & 0.007         & 0.002 & 0.021 \\
Signal modeling                                                       & 0.018         & 0.003         & 0.021         & 0.017 & 0.019 \\
Signal shape and self cross-feed                                      & 0.002         & 0.002         & 0.002         & 0.002 & 0.002 \\ \hline
Total                                                                 & 0.027         & 0.052         & 0.023         & 0.017 & 0.035 \\ \hline \hline
\end{tabular}
\end{table*}

\section{FORWARD-BACKWARD ASYMMETRY}

\begin{figure*}[htb]
\centering
\subfigure[$B \rightarrow X_s e^+ e^-$     candidates with $\cos\theta > 0$]{
\includegraphics*[width=7cm]{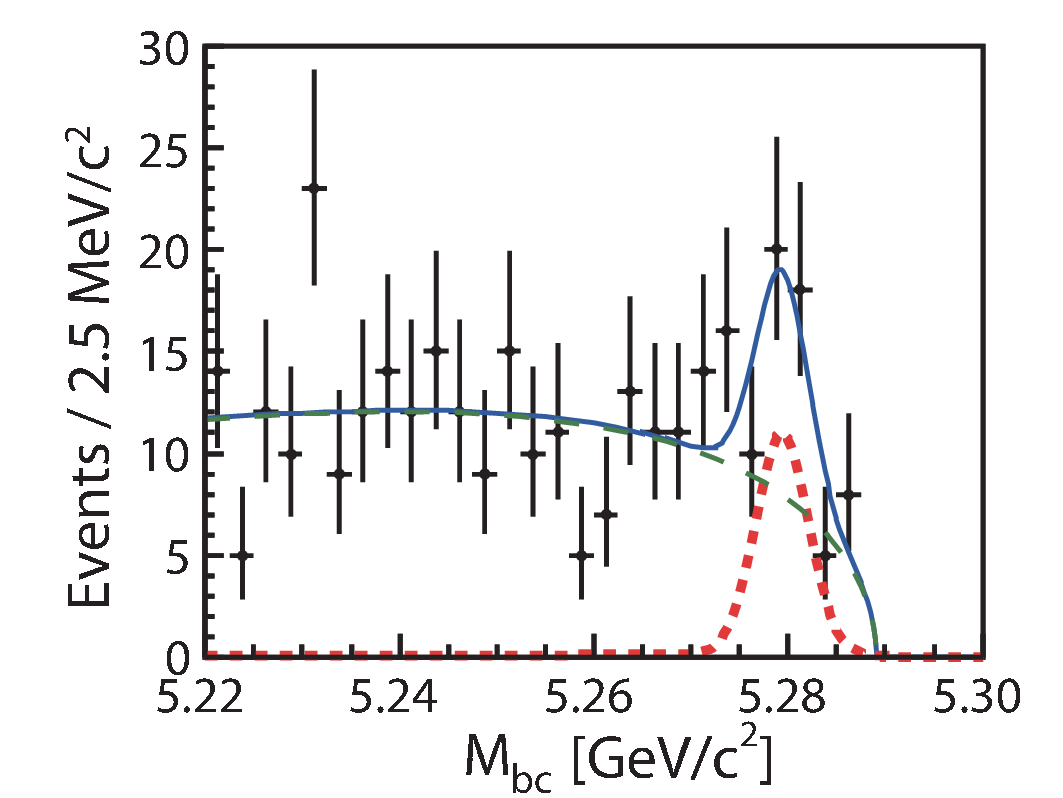}
\label{fig:mbc_1st_xsee_forward}}
\subfigure[$B \rightarrow X_s e^+ e^-$     candidates with $\cos\theta < 0$]{
\includegraphics*[width=7cm]{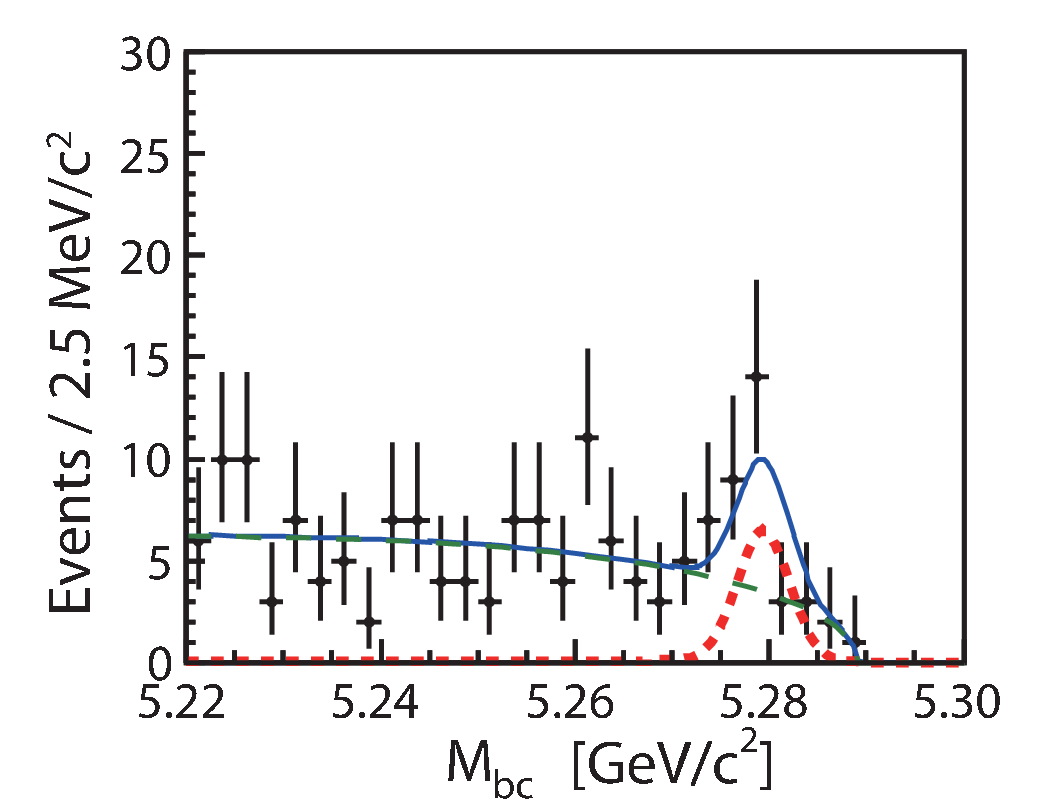}
\label{fig:mbc_1st_xsee_backward}}
\subfigure[$B \rightarrow X_s \mu^+ \mu^-$     candidates with $\cos\theta > 0$]{
\includegraphics*[width=7cm]{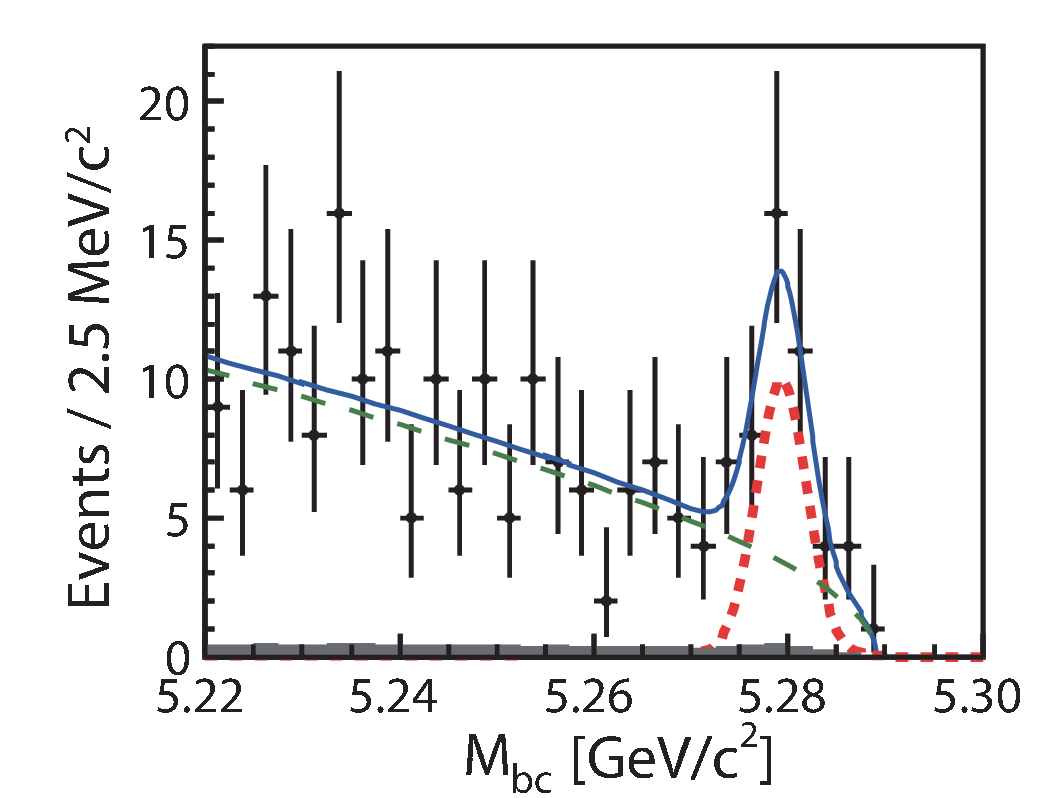}
\label{fig:mbc_1st_xsmm_forward}}
\subfigure[$B \rightarrow X_s \mu^+ \mu^-$     candidates with $\cos\theta < 0$]{
\includegraphics*[width=7cm]{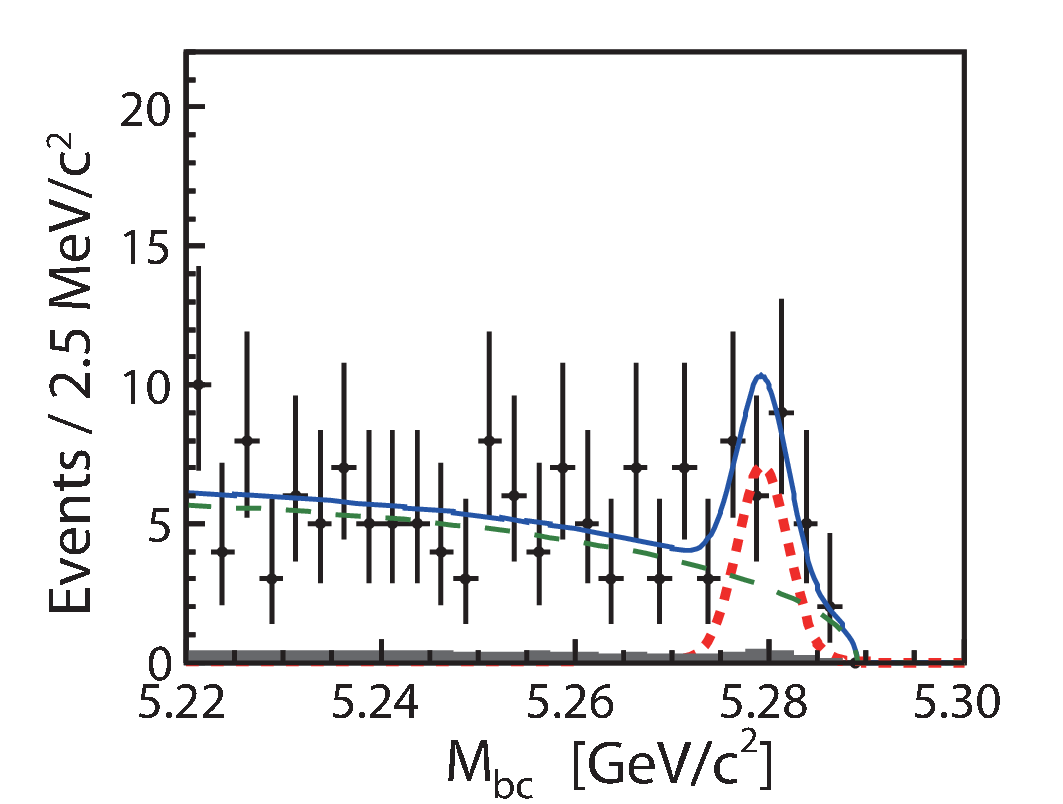}
\label{fig:mbc_1st_xsmm_backward}}
\caption{
$M_{\rm bc}$ distributions in 1st $q^2$ bin for
(a) $B \rightarrow X_s e^+ e^-$     candidates with $\cos\theta > 0$,
(b) $B \rightarrow X_s e^+ e^-$     candidates with $\cos\theta < 0$,
(c) $B \rightarrow X_s \mu^+ \mu^-$ candidates with $\cos\theta > 0$, and
(d) $B \rightarrow X_s \mu^+ \mu^-$ candidates with $\cos\theta < 0$.
The thicker dashed curve (red) shows the sum of the signal
and the self cross-feed components.
The thinner dashed curve (green) shows the combinatorial background component.
The filled histogram (gray) shows the peaking background component.
The sums of all components are shown by the solid curve (blue).
\label{fig:mbc_1st}
}
\end{figure*}

\begin{figure*}[htb]
\centering
\subfigure[$B \rightarrow X_s e^+ e^-$     candidates with $\cos\theta > 0$]{
\includegraphics*[width=7cm]{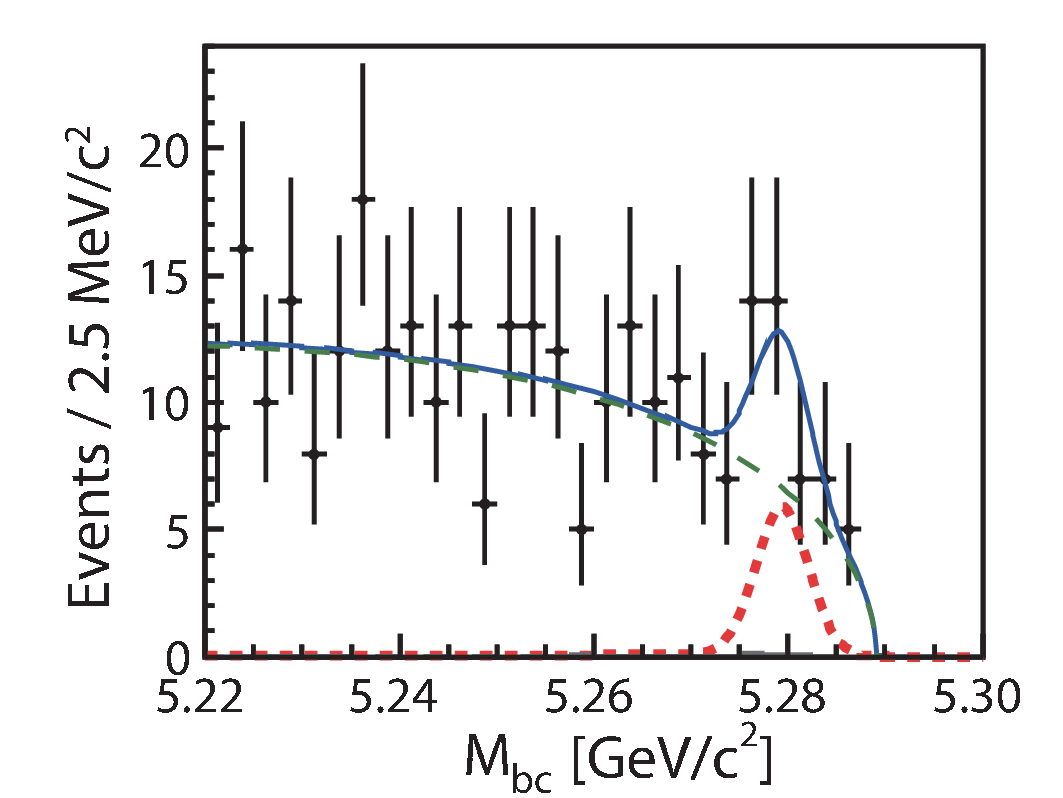}
\label{fig:mbc_2nd_xsee_forward}}
\subfigure[$B \rightarrow X_s e^+ e^-$     candidates with $\cos\theta < 0$]{
\includegraphics*[width=7cm]{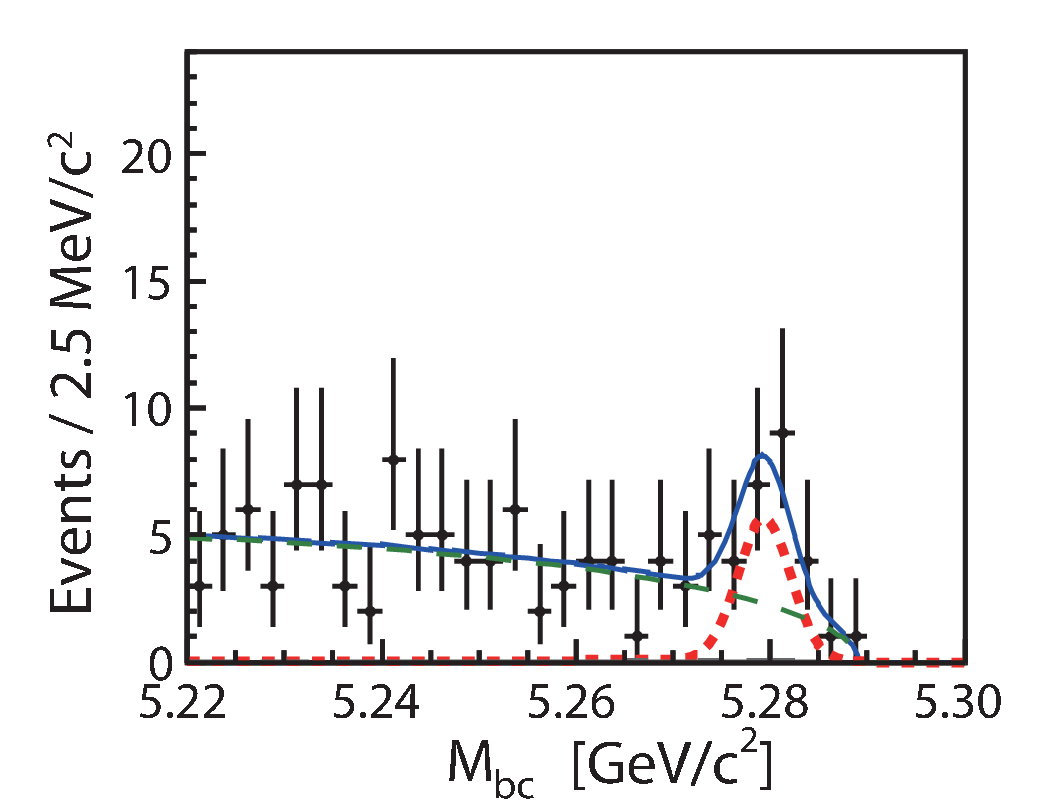}
\label{fig:mbc_2nd_xsee_backward}}
\subfigure[$B \rightarrow X_s \mu^+ \mu^-$     candidates with $\cos\theta > 0$]{
\includegraphics*[width=7cm]{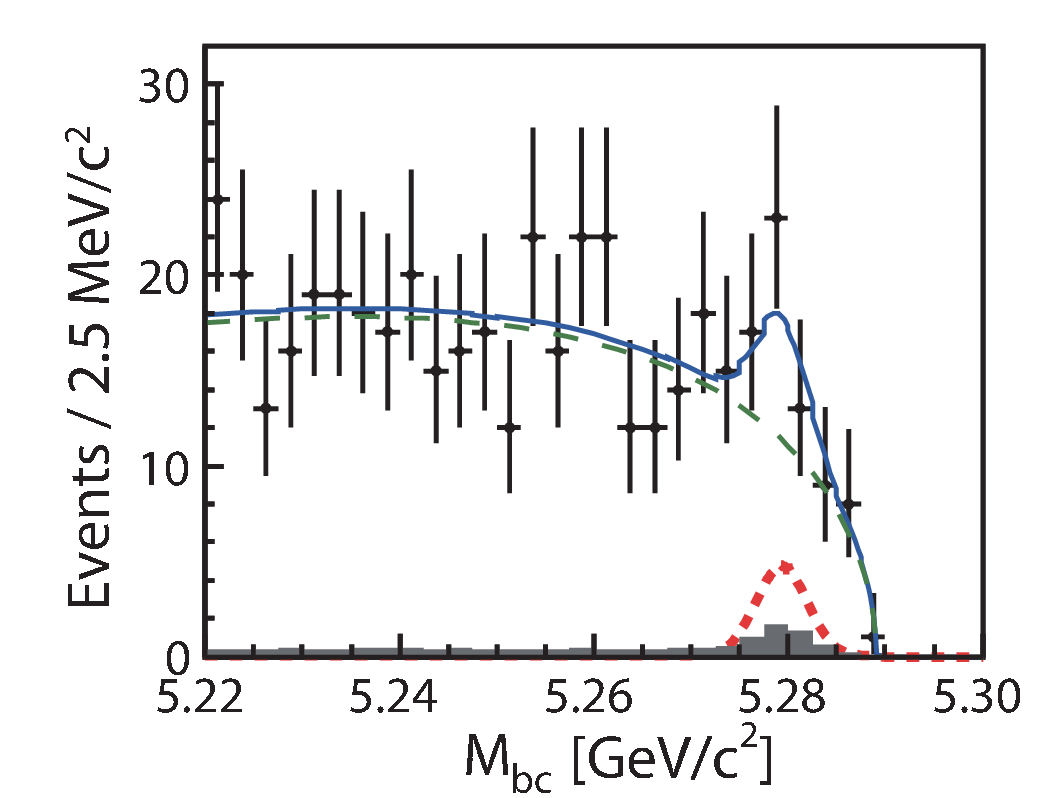}
\label{fig:mbc_2nd_xsmm_forward}}
\subfigure[$B \rightarrow X_s \mu^+ \mu^-$     candidates with $\cos\theta < 0$]{
\includegraphics*[width=7cm]{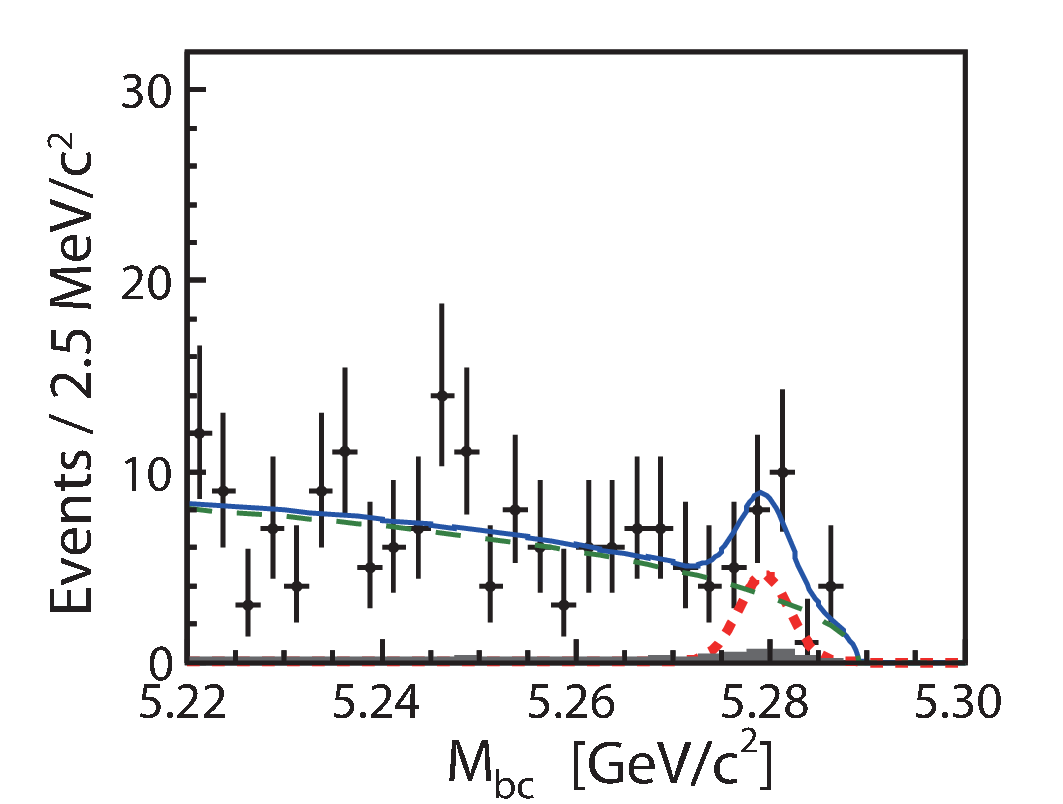}
\label{fig:mbc_2nd_xsmm_backward}}
\caption{
$M_{\rm bc}$ distributions in 2nd $q^2$ bin for
(a) $B \rightarrow X_s e^+ e^-$     candidates with $\cos\theta > 0$,
(b) $B \rightarrow X_s e^+ e^-$     candidates with $\cos\theta < 0$,
(c) $B \rightarrow X_s \mu^+ \mu^-$ candidates with $\cos\theta > 0$, and
(d) $B \rightarrow X_s \mu^+ \mu^-$ candidates with $\cos\theta < 0$.
The thicker dashed curve (red) shows the sum of the signal
and the self cross-feed components.
The thinner dashed curve (green) shows the combinatorial background component.
The filled histogram (gray) shows the peaking background component.
The sums of all components are shown by the solid curve (blue).
\label{fig:mbc_2nd}
}
\end{figure*}

\begin{figure*}[htb]
\centering
\subfigure[$B \rightarrow X_s e^+ e^-$     candidates with $\cos\theta > 0$]{
\includegraphics*[width=7cm]{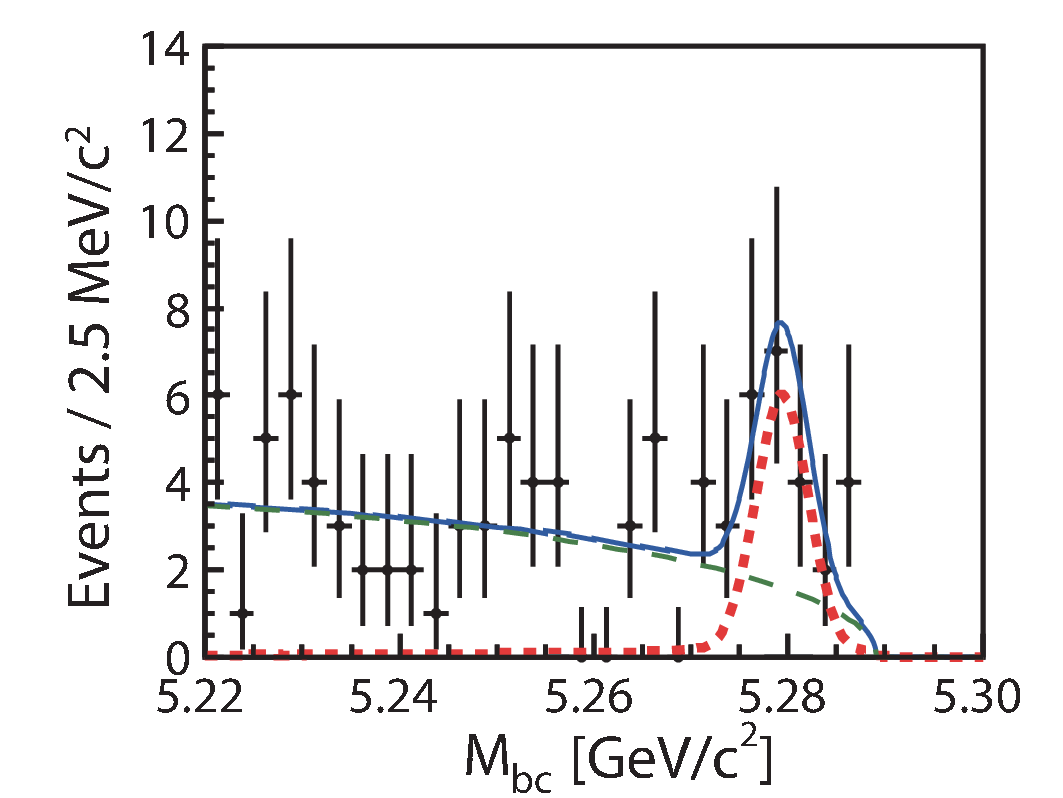}
\label{fig:mbc_3rd_xsee_forward}}
\subfigure[$B \rightarrow X_s e^+ e^-$     candidates with $\cos\theta < 0$]{
\includegraphics*[width=7cm]{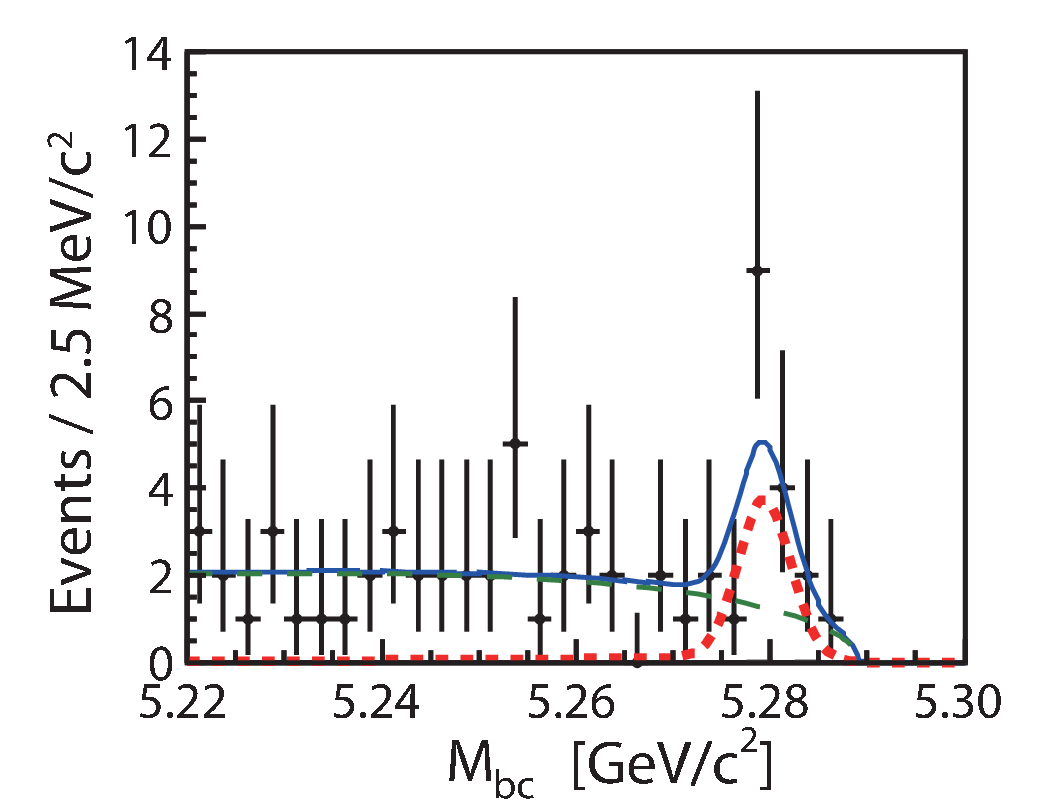}
\label{fig:mbc_3rd_xsee_backward}}
\subfigure[$B \rightarrow X_s \mu^+ \mu^-$     candidates with $\cos\theta > 0$]{
\includegraphics*[width=7cm]{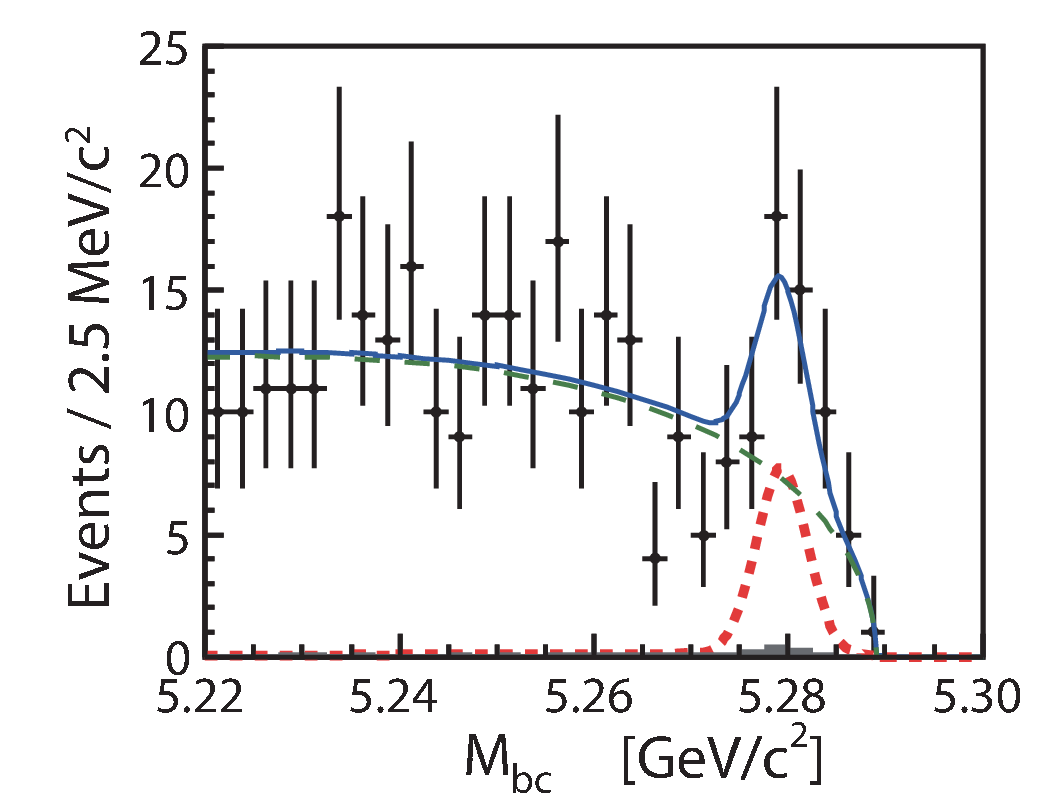}
\label{fig:mbc_3rd_xsmm_forward}}
\subfigure[$B \rightarrow X_s \mu^+ \mu^-$     candidates with $\cos\theta < 0$]{
\includegraphics*[width=7cm]{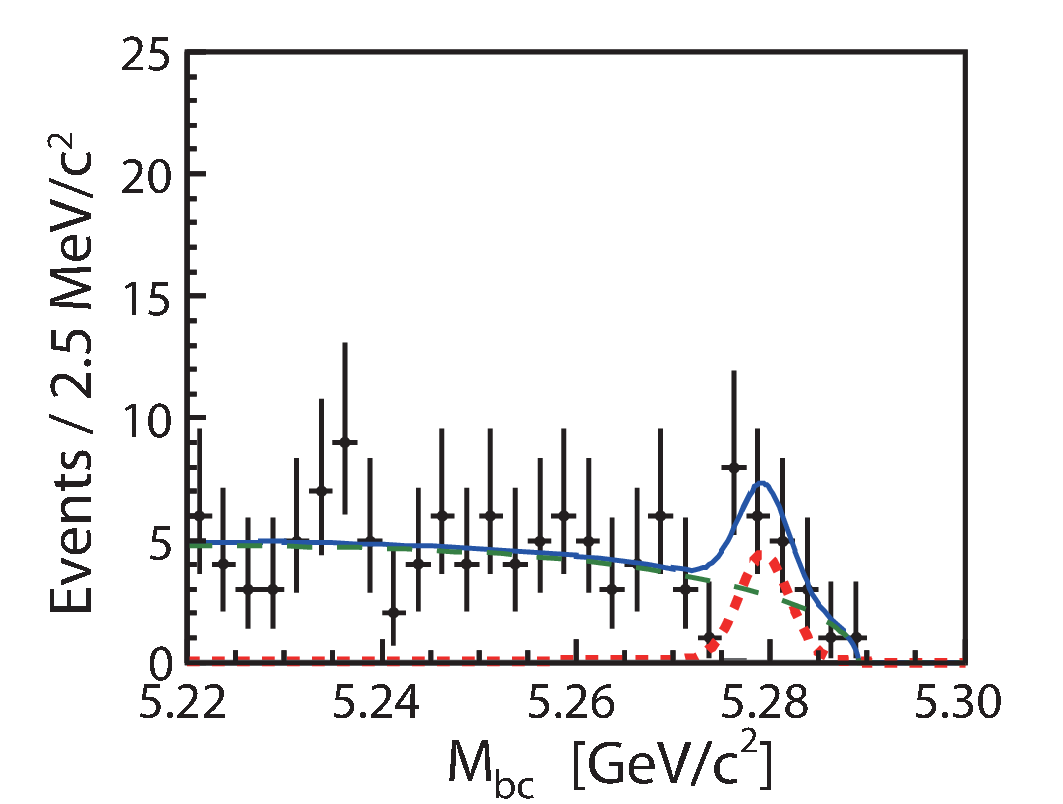}
\label{fig:mbc_3rd_xsmm_backward}}
\caption{
$M_{\rm bc}$ distributions in 3rd $q^2$ bin for
(a) $B \rightarrow X_s e^+ e^-$     candidates with $\cos\theta > 0$,
(b) $B \rightarrow X_s e^+ e^-$     candidates with $\cos\theta < 0$,
(c) $B \rightarrow X_s \mu^+ \mu^-$ candidates with $\cos\theta > 0$, and
(d) $B \rightarrow X_s \mu^+ \mu^-$ candidates with $\cos\theta < 0$.
The thicker dashed curve (red) shows the sum of the signal
and the self cross-feed components.
The thinner dashed curve (green) shows the combinatorial background component.
The filled histogram (gray) shows the peaking background component.
The sums of all components are shown by the solid curve (blue).
\label{fig:mbc_3rd}
}
\end{figure*}

\begin{figure*}[htb]
\centering
\subfigure[$B \rightarrow X_s e^+ e^-$     candidates with $\cos\theta > 0$]{
\includegraphics*[width=7cm]{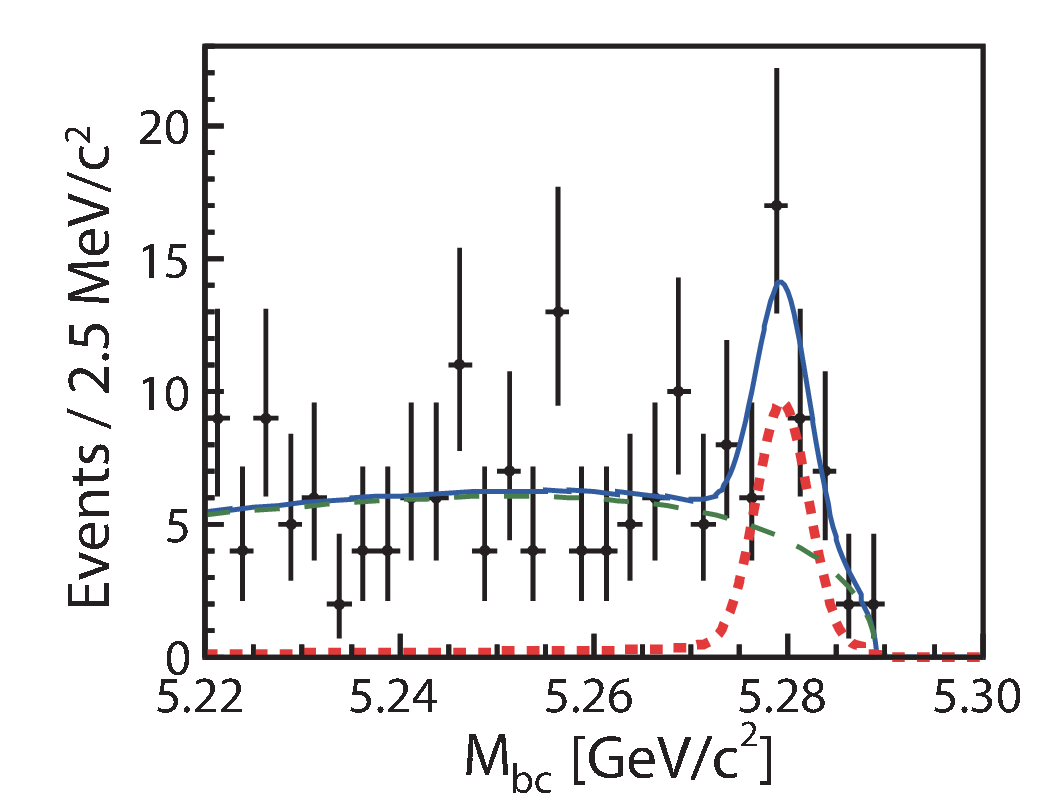}
\label{fig:mbc_4th_xsee_forward}}
\subfigure[$B \rightarrow X_s e^+ e^-$     candidates with $\cos\theta < 0$]{
\includegraphics*[width=7cm]{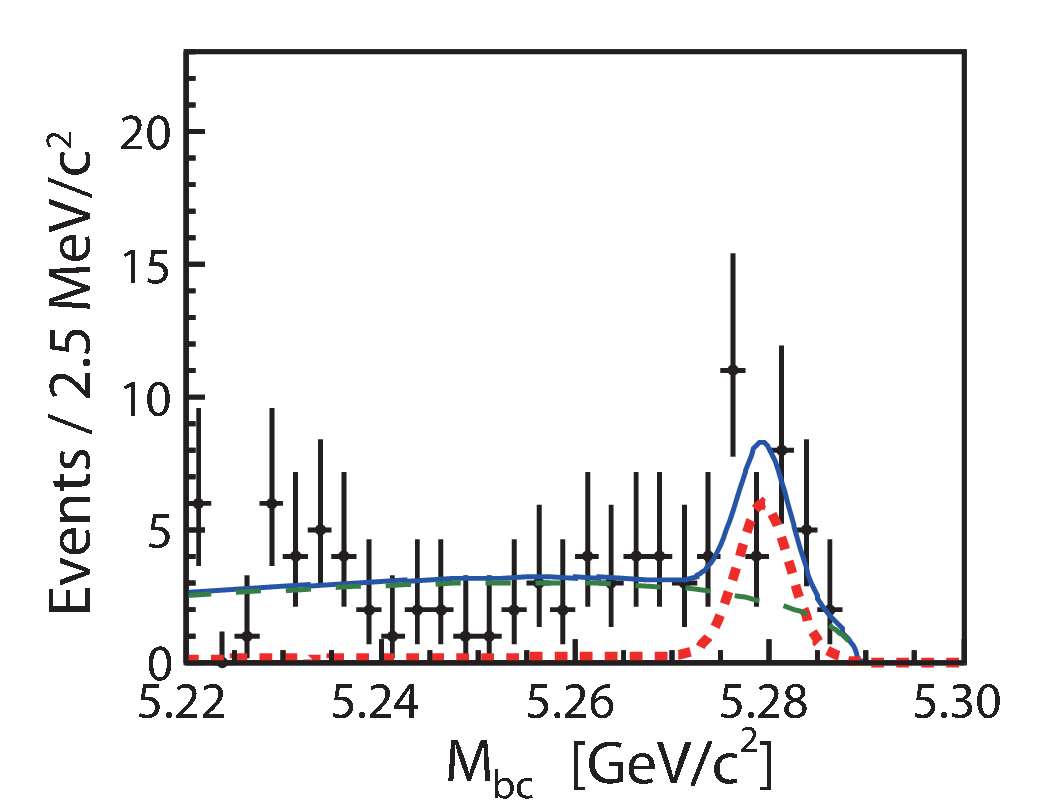}
\label{fig:mbc_4th_xsee_backward}}
\subfigure[$B \rightarrow X_s \mu^+ \mu^-$     candidates with $\cos\theta > 0$]{
\includegraphics*[width=7cm]{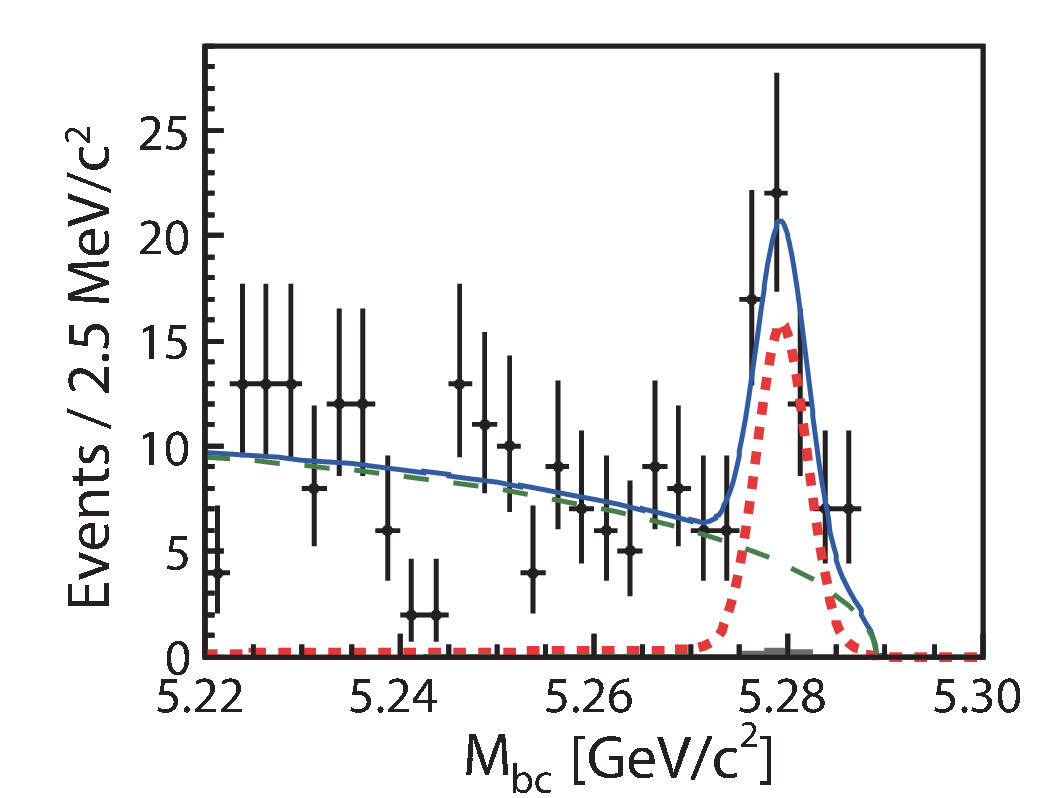}
\label{fig:mbc_4th_xsmm_forward}}
\subfigure[$B \rightarrow X_s \mu^+ \mu^-$     candidates with $\cos\theta < 0$]{
\includegraphics*[width=7cm]{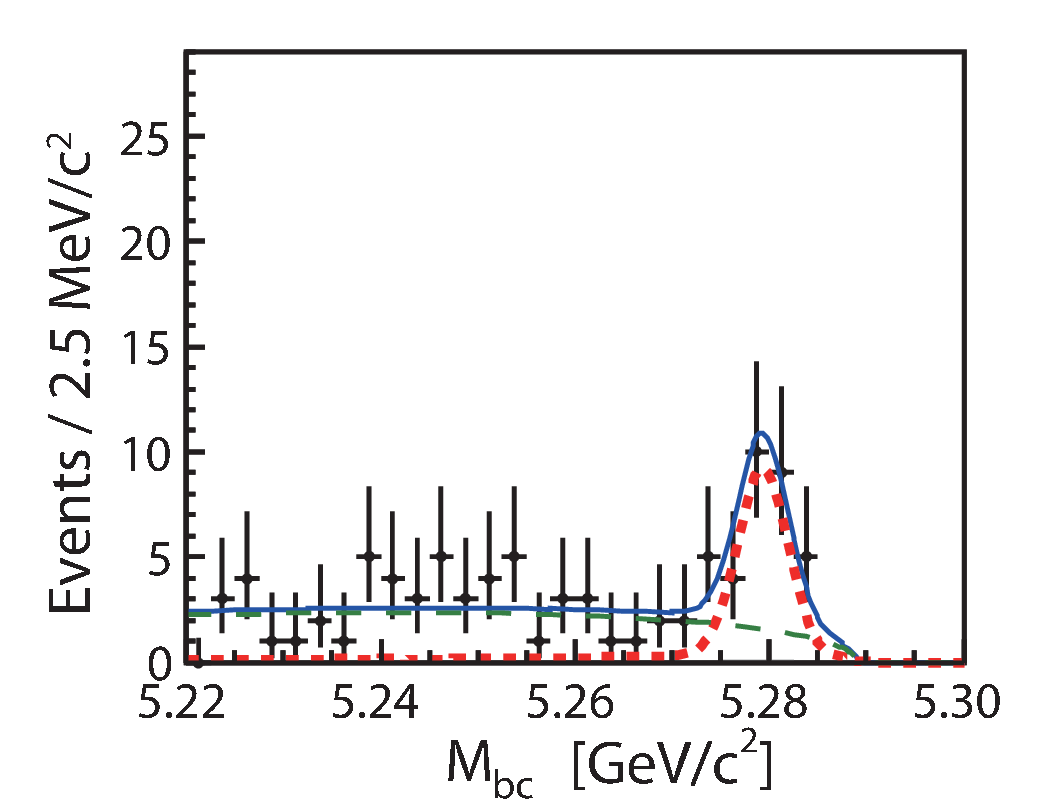}
\label{fig:mbc_4th_xsmm_backward}}
\caption{
$M_{\rm bc}$ distributions in 4th $q^2$ bin for
(a) $B \rightarrow X_s e^+ e^-$     candidates with $\cos\theta > 0$,
(b) $B \rightarrow X_s e^+ e^-$     candidates with $\cos\theta < 0$,
(c) $B \rightarrow X_s \mu^+ \mu^-$ candidates with $\cos\theta > 0$, and
(d) $B \rightarrow X_s \mu^+ \mu^-$ candidates with $\cos\theta < 0$.
The thicker dashed curve (red) shows the sum of the signal
and the self cross-feed components.
The thinner dashed curve (green) shows the combinatorial background component.
The filled histogram (gray) shows the peaking background component.
The sums of all components are shown by the solid curve (blue).
\label{fig:mbc_4th}
}
\end{figure*}

\begin{figure*}[htb]
\centering
\subfigure[$B \rightarrow X_s e^+ e^-$     candidates with $\cos\theta > 0$]{
\includegraphics*[width=7cm]{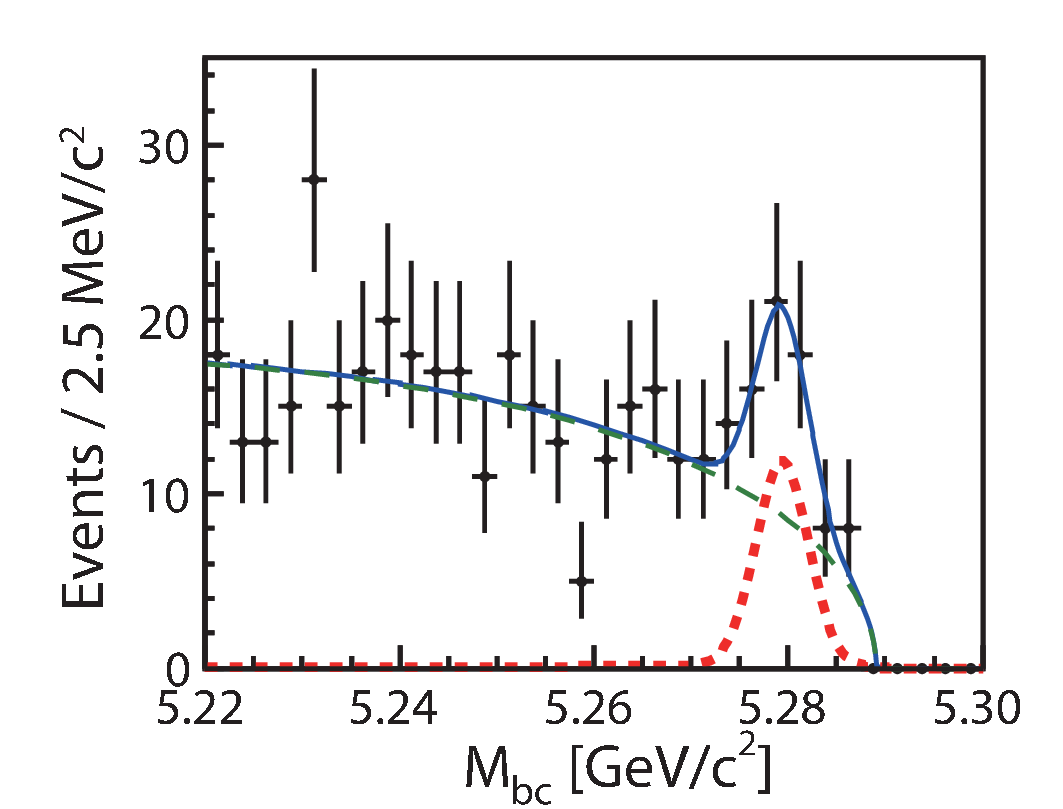}
\label{fig:mbc_16q2_xsee_forward}}
\subfigure[$B \rightarrow X_s e^+ e^-$     candidates with $\cos\theta < 0$]{
\includegraphics*[width=7cm]{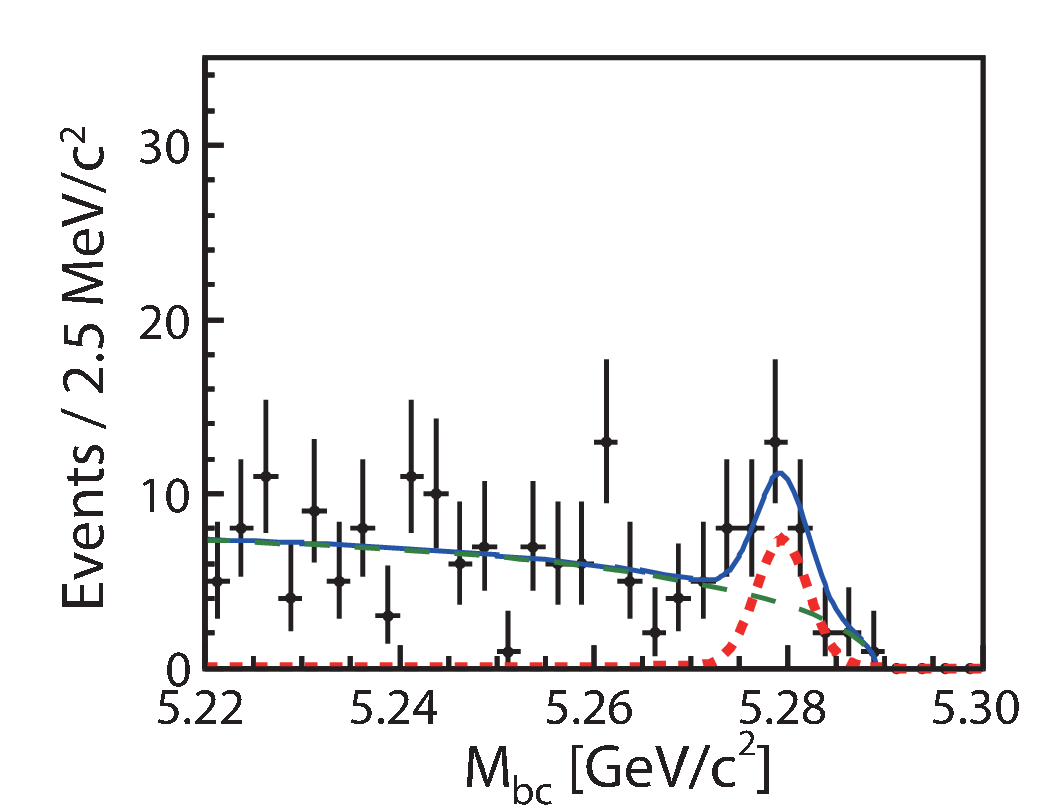}
\label{fig:mbc_16q2_xsee_backward}}
\subfigure[$B \rightarrow X_s \mu^+ \mu^-$     candidates with $\cos\theta > 0$]{
\includegraphics*[width=7cm]{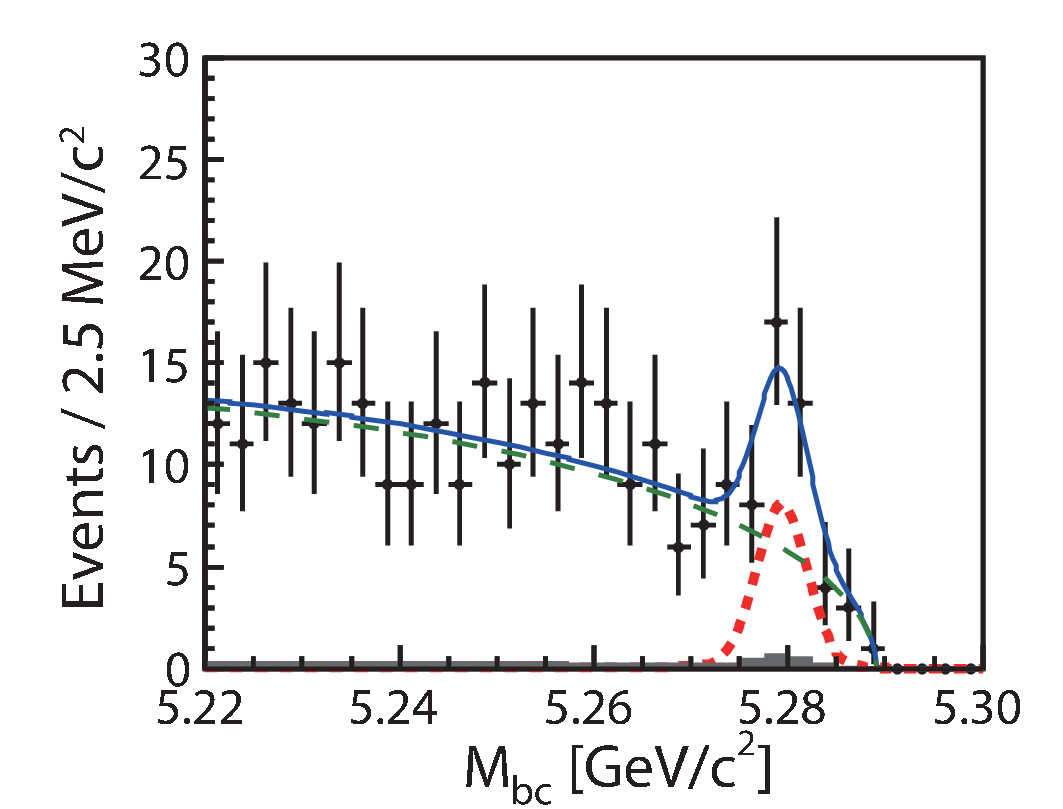}
\label{fig:mbc_16q2_xsmm_forward}}
\subfigure[$B \rightarrow X_s \mu^+ \mu^-$     candidates with $\cos\theta < 0$]{
\includegraphics*[width=7cm]{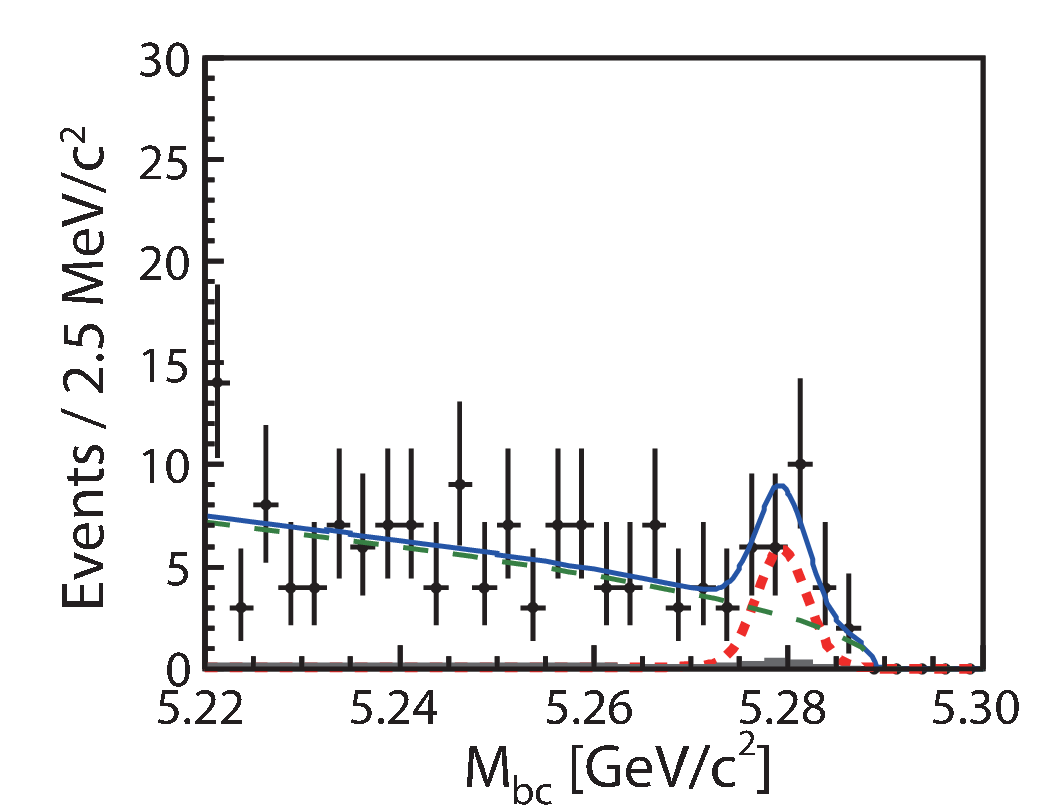}
\label{fig:mbc_16q2_xsmm_backward}}
\caption{
$M_{\rm bc}$ distributions in the low-$q^2$ region, $1 < q^2 < 6$ GeV$^2$ for
(a) $B \rightarrow X_s e^+ e^-$     candidates with $\cos\theta > 0$,
(b) $B \rightarrow X_s e^+ e^-$     candidates with $\cos\theta < 0$,
(c) $B \rightarrow X_s \mu^+ \mu^-$ candidates with $\cos\theta > 0$, and
(d) $B \rightarrow X_s \mu^+ \mu^-$ candidates with $\cos\theta < 0$.
The thicker dashed curve (red) shows the sum of the signal
and the self cross-feed components.
The thinner dashed curve (green) shows the combinatorial background component.
The filled histogram (gray) shows the peaking background component.
The sums of all components are shown by the solid curve (blue).
\label{fig:mbc_16q2}
}
\end{figure*}

\begin{figure}[hb]
{\includegraphics[width=10.0cm]{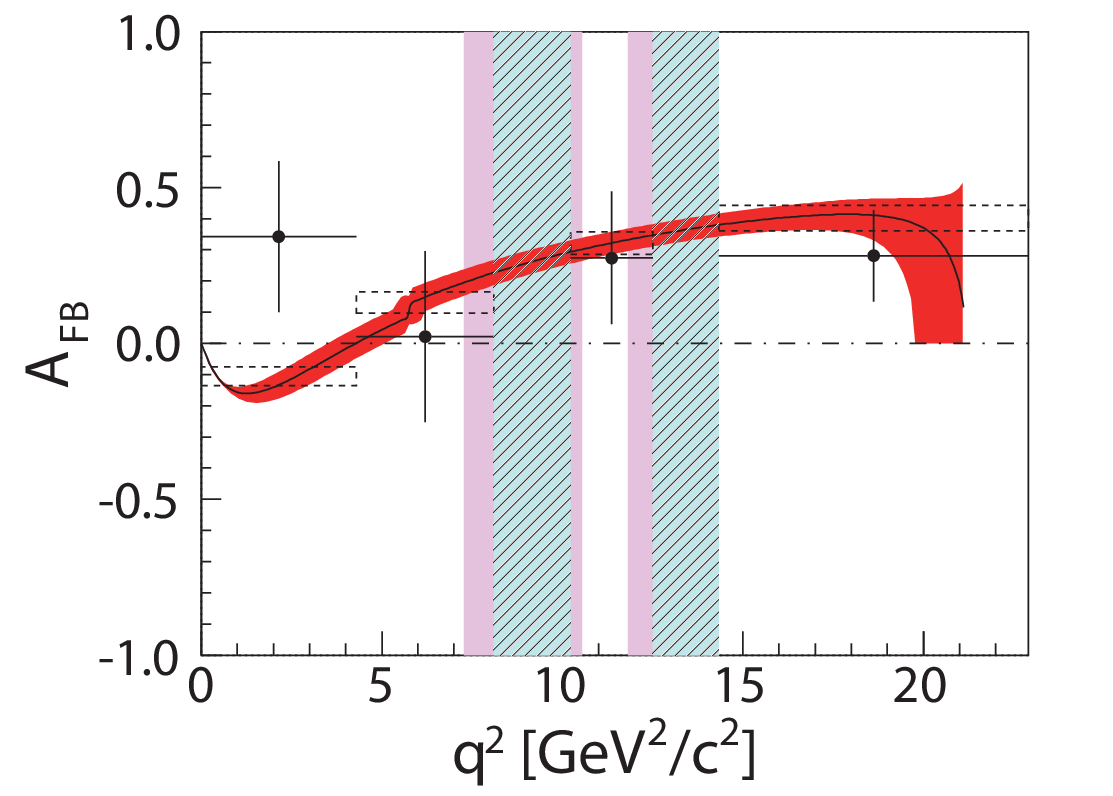}}
\caption{
Measured ${\cal A}_{\rm FB}$ as a function of $q^2$.
The curve (black) with the band (red) and dashed boxes (black) represent the SM prediction 
while filled circles with error bars show the fit results.
The $J/\psi$ and $\psi$(2S) veto regions are shown as teal hatched regions.
For the electron channel, the pink shaded regions are added to the veto regions
due to the large bremsstrahlung effect.
The uncertainty on the SM prediction is estimated
by varying
the $b$-quark mass ($4.80 \pm 0.15$ GeV/$c^2$),
the $s$-quark mass ($0.20 \pm 0.10$ GeV/$c^2$),
and the renormalization scale ($\mu$ = 2.5 and 5 GeV)
\cite{THEORY_XSLL, IMDEP_THEORY}.
The lower edge of the uncertainty is set to zero
in the $q^2$ region larger than maximum possible value,
which is determined by the masses of the bottom and strange quarks.
}
\label{fig:q2_afb}
\end{figure}

Figure \ref{fig:mbc_1st}, \ref{fig:mbc_2nd}, \ref{fig:mbc_3rd}, \ref{fig:mbc_4th}, and \ref{fig:mbc_16q2} show
the $M_{\rm bc}$ distributions for
$B \rightarrow X_s e^+ e^-$ and $B \rightarrow X_s \mu^+ \mu^-$ candidates
with positive and negative $\cos\theta$ in each $q^2$ bin.
The total signal yields for
$B \rightarrow X_s e^+ e^-$ and
$B \rightarrow X_s \mu^+ \mu^-$
are 140 $\pm$ 19(stat)
and 161 $\pm$ 20(stat), respectively. 
The fit results obtained in each $q^2$ bins
are summarized in Table~\ref{tab:results}.
Figure \ref{fig:q2_afb} shows the ${\cal A}_{\rm FB}$ distribution
as a function of $q^2$.
The ${\cal A}_{\rm FB}$ results are found to be consistent with the SM prediction
in the 2nd to 4th $q^2$ bins,
while it deviates from the SM
in the 1st $q^2$ bin by 1.8$\sigma$;
here, the systematic uncertainty
is taken into account.
The results in the 3rd and 4th bin also excludes
${\cal A}_{\rm FB} < 0$ at the 2.3$\sigma$ level.

To distinguish the contributions from $B \rightarrow K \ell^+ \ell^-$,
$B \rightarrow K^* \ell^+ \ell^-$, and non-$K^{(*)} \ell^+ \ell^-$ candidates,
we divide the samples into distinct $M_{X_s}$ ranges
and extract ${\cal A}_{\rm FB}$ by the same fitting method.
Table~\ref{tab:afb_separate} shows the ${\cal A}_{\rm FB}$ values
in each subsample.
${\cal A}_{\rm FB}$ in $B \rightarrow K \ell^+ \ell^-$ is consistent with null,
as expected in the SM,
while ${\cal A}_{\rm FB}$ in $B \rightarrow K^* \ell^+ \ell^-$ is consistent with
previous measurements \cite{AFB_XSLL_BELLE,AFB_XSLL_BABAR,AFB_XSLL_CDF,AFB_XSLL_LHCb,AFB_XSLL_CMS}.

\begin{table*}[htb]
\caption{
Fit results for subsamples of 
(i) $B \rightarrow K \ell^+ \ell^-$,
(ii) $B \rightarrow K^- \pi^+ \ell^+ \ell^-$, $K^- \pi^0 \ell^+ \ell^-$, or $K_S^0 \pi^- \ell^+ \ell^-$ with $M_{X_s} < 1.1$ GeV/$c^2$,
and
(iii) $B \rightarrow X_s \ell^+ \ell^-$ with $M_{X_s} > 1.1$ GeV/$c^2$
for the five $q^2$ bins.
The uncertainty includes only statistical uncertainty.
Unfortunately, ${\cal A}_{\rm FB}$ for $B \rightarrow K \ell^+ \ell^-$ can not be obtained in 3rd $q^2$ bin,
due to too low statistics.
}
\label{tab:afb_separate}
\begin{tabular}
{c|ccccc}
\hline \hline
State & \hspace{1.0em}1st $q^2$ bin & \hspace{1.0em}2nd $q^2$ bin & \hspace{1.0em}3rd $q^2$ bin & \hspace{1.0em}4th $q^2$ bin & \hspace{1.0em}$1 < q^2 < 6$ GeV$^2$/$c^2$ \\ \hline
$K$                                  & $-0.05 \pm 0.24$ & $-0.11 \pm 0.29$ &   n.a.          & $ 0.12 \pm 0.18$ & $0.00 \pm 0.13$\\
$K^*$ with $M_{X_s} < 1.1$ GeV/$c^2$ & $ 0.62 \pm 0.42$ & $ 0.20 \pm 0.33$ & $0.01 \pm 0.34$ & $ 0.21 \pm 0.22$ & $0.55 \pm 0.43$\\
$X_s$ with $M_{X_s} > 1.1$ GeV/$c^2$ & $ 0.25 \pm 0.45$ & $ 0.97 \pm 0.60$ & $0.92 \pm 0.32$ & $ 0.65 \pm 0.54$ & $0.74 \pm 0.54$\\
\hline \hline
\end{tabular}
\end{table*}

\section{CONCLUSION}
In conclusion,
we report the first measurement of
the lepton forward-backward asymmetry
for the electroweak penguin process
$B \rightarrow X_s \ell^+ \ell^-$
using a data sample containing $772\times10^6$ $B\bar{B}$ pairs
collected with the Belle detector.
${\cal A}_{\rm FB}$ for the inclusive $B \rightarrow X_s \ell^+ \ell^-$ is extrapolated from the sum of 10 exclusive $X_s$ states,
assuming ${\cal A}_{\rm FB}$ depends neither on the lepton flavor nor on the $X_s$ mass.
For $q^2 > 10.2$ GeV$^2$/$c^2$,
${\cal A}_{\rm FB} < 0$ is excluded at the $2.3\sigma$ level.
For $q^2 < 4.3$ GeV$^2$/$c^2$,
the result is within 1.8$\sigma$ of the SM expectation.
The results can be used to constrain
various extensions of the SM.

\section{ACKNOWLEDGMENT}
We thank T.~Morozumi and T.~Goto for their invaluable suggestions.
We thank the KEKB group for the excellent operation of the
accelerator; the KEK cryogenics group for the efficient
operation of the solenoid; and the KEK computer group,
the National Institute of Informatics, and the 
PNNL/EMSL computing group for valuable computing
and SINET4 network support.  We acknowledge support from
the Ministry of Education, Culture, Sports, Science, and
Technology (MEXT) of Japan, the Japan Society for the 
Promotion of Science (JSPS), and the Tau-Lepton Physics 
Research Center of Nagoya University; 
the Australian Research Council;
Austrian Science Fund under Grant No.~P 22742-N16 and P 26794-N20;
the National Natural Science Foundation of China under Contracts 
No.~10575109, No.~10775142, No.~10875115, No.~11175187, No.~11475187
and No.~11575017;
the Chinese Academy of Science Center for Excellence in Particle Physics; 
the Ministry of Education, Youth and Sports of the Czech
Republic under Contract No.~LG14034;
the Carl Zeiss Foundation, the Deutsche Forschungsgemeinschaft, the
Excellence Cluster Universe, and the VolkswagenStiftung;
the Department of Science and Technology of India; 
the Istituto Nazionale di Fisica Nucleare of Italy; 
the WCU program of the Ministry of Education, National Research Foundation (NRF) 
of Korea Grants No.~2011-0029457,  No.~2012-0008143,  
No.~2012R1A1A2008330, No.~2013R1A1A3007772, No.~2014R1A2A2A01005286, 
No.~2014R1A2A2A01002734, No.~2015R1A2A2A01003280 , No. 2015H1A2A1033649;
the Basic Research Lab program under NRF Grant No.~KRF-2011-0020333,
Center for Korean J-PARC Users, No.~NRF-2013K1A3A7A06056592; 
the Brain Korea 21-Plus program and Radiation Science Research Institute;
the Polish Ministry of Science and Higher Education and 
the National Science Center;
the Ministry of Education and Science of the Russian Federation and
the Russian Foundation for Basic Research;
the Slovenian Research Agency;
Ikerbasque, Basque Foundation for Science and
the Euskal Herriko Unibertsitatea (UPV/EHU) under program UFI 11/55 (Spain);
the Swiss National Science Foundation; 
the Ministry of Education and the Ministry of Science and Technology of Taiwan;
and the U.S.\ Department of Energy and the National Science Foundation.
This work is supported by a Grant-in-Aid from MEXT for 
Science Research in a Priority Area (``New Development of 
Flavor Physics'') and from JSPS for Creative Scientific 
Research (``Evolution of Tau-lepton Physics'').


\begin{thebibliography}{99}

\bibitem{CHARGE_CONJUGATION}
Charge-conjugate decays are implied throughout this paper,
unless otherwise stated.

\bibitem{OPE}
K.~G.~Wilson, Phys. Rev. {\bf 179}, 1499 (1969).

\bibitem{WC_XSLL}
G.~Buchalla, A.J.~Buras, and M.E.~Lautenbacher,
Rev. Mod. Phys. {\bf 68}, 1125 (1996).

\bibitem{THEORY_XSLL}
A.~Ali, E.~Lunghi, C.~Greub, and G.~Hiller,
Phys. Rev. {\bf D66}, 034002 (2002).

\bibitem{BR_XSLL_BELLE}
M.~Iwasaki {\it et al.}
(Belle Collaboration),
Phys. Rev. {\bf D72}, 092005 (2005).

\bibitem{BR_XSLL_BABAR}
B.~Aubert {\it et al.}
(\mbox{\sl B\hspace{-0.4em} {\small\sl A}\hspace{-0.37em} \sl B\hspace{-0.4em}
{\small\sl A\hspace{-0.02em}R}}
Collaboration),
Phys. Rev. Lett. {\bf 93}, 081802 (2004).

\bibitem{IMDEP_THEORY}
S.~Fukae, C.S.~Kim, T.~Morozumi, and T.~Yoshikawa,
Phys. Rev. {\bf D59}, 074013 (1999).

\bibitem{IMDEP_THEORY_NEW}
T.~Huber, T.~Hurth, and E.~Lunghi,
Nucl. Phys. {\bf B802}, 40 (2008).

\bibitem{AFB_XSLL_BELLE}
J.-T.~Wei {\it et al.}
(Belle Collaboration),
Phys. Rev. Lett. {\bf 103}, 171801 (2009).

\bibitem{AFB_XSLL_BABAR}
B.~Aubert, {\it et al.}
(\mbox{\sl B\hspace{-0.4em} {\small\sl A}\hspace{-0.37em} \sl B\hspace{-0.4em}
{\small\sl A\hspace{-0.02em}R}}
Collaboration),
Phys. Rev. {\bf D79}, 031102(R) (2009).

\bibitem{AFB_XSLL_CDF}
T.~Aaltonen {\it et al.}
(CDF Collaboration),
Phys. Rev. Lett. {\bf 108}, 081807 (2012).

\bibitem{AFB_XSLL_LHCb}
R.~Aaij {\it et al.}
(LHCb Collaboration),
arXiv:1512.04442.

\bibitem{AFB_XSLL_CMS}
S.~Chatrchyan {\it et al.} (CMS Collaboration),
Phys. Lett. {\bf B727}, 77 (2013).

\bibitem{AFB_XSLL_EQ}
K.S.M.Lee, Z~Ligeti, I.W.~Stewart,
and Frank~J.~Tackmann,
Phys. Rev. {\bf D75}, 034016 (2007).

\bibitem{CKM}
M.~Kobayashi and T.~Maskawa, Prog. Theor. Phys. {\bf 49}, 652 (1973); 
N.~Cabibbo, Phys. Rev. Lett. 10, 531 (1963).

\bibitem{BELLE}
A.~Abashian {\it et al.} (Belle Collaboration),
Nucl. Instrum. Meth. A {\bf 479}, 117 (2002);
also see the detector section in J.~Brodzicka {\it et al.}, Prog. Theor. Exp. Phys. (2012) 04D001.

\bibitem{KEKB}
S.~Kurokawa and E.~Kikutani,
Nucl. Instrum. Meth. A {\bf 499}, 1 (2003),
and other papers included in this volume;
T.~Abe {\it et al.}, Prog. Theor. Exp. Phys. (2013) 03A001
and following articles up to 03A011.
 
\bibitem{EVTGEN_KLL}
A.~Ali, P.~Ball, L.T.~Handoko, and G.~Hiller,
Phys. Rev. {\bf D61}, 074024 (2000).

\bibitem{EVTGEN_XSLL}
F.~Kruger and L.M.~Sehgal,
Phys. Lett. {\bf B380}, 199 (1996).

\bibitem{FERMIMOTION}
A.~Ali and E.~Pietarinen,
Nucl. Phys. {\bf B154}, 519 (1979).


\bibitem{HFAG}
Y.~Amhis {\it et al.}, arXiv:1207.1158
and online update at
http://www.slac.stanford.edu/xorg/hfag/

\bibitem{PID}
E.~Nakano,
Nucl. Instrum. Meth. A {\bf 494}, 402 (2002).

\bibitem{NEUROBAYES}
M.~Feindt and U.~Kerzel,
Nucl. Instrum. Meth. A {\bf 559}, 190 (2006).

\bibitem{EVISMMISS}
We define the total visible energy
$E_{\rm vis} = \sum_i E_i^*$,
and the missing mass
$M_{\rm miss} = \sqrt{(2E_{\rm beam}^* - \sum_i E_i^*)^2 - \sum_i |\vec{p}_i^*|^2}$,
where $(\vec{p}_i^*, E_i^*)$ are the reconstructed four-momenta
in the $\Upsilon(4S)$ rest frame
of all tracks assumed to be pions
and all photons in the event.

\bibitem{Fox}
The Fox-Wolfram moments were introduced in
G.C.~Fox and S.~Wolfram,
Phys. Rev. Lett. {\bf 41}, 1581 (1978);
The modified moments used in this Letter are 
described in
S.H.~Lee {\it et al.}
(Belle Collaboration),
Phys. Rev. Lett. {\bf 91}, 261801 (2003).

\bibitem{ARGUS} 
H.~Albrecht {\it et al.} (ARGUS Collaboration),
Phys. Lett. B {\bf 241}, 278 (1990). 

\bibitem{SYS_FF1}
P.~Ball and R.~Zwicky,
Phys. Rev. {\bf D71}, 014015 (2005).

\bibitem{SYS_FF2}
P.~Ball and R.~Zwicky,
Phys. Rev. {\bf D71}, 014029 (2005).

\bibitem{SYS_FERMI_SEMI}
C.~Schwanda {\it et al.}
(Belle Collaboration),
Phys. Rev. {\bf D75}, 032005 (2007).

\bibitem{SYS_FERMI_XSGAMMA}
A.~Limosani {\it et al.}
(Belle Collaboration),
Phys. Rev. Lett. {\bf 103}, 241801 (2009).

\bibitem{SWAVE}
R.~Aaij {\it et al.}
(LHCb Collaboration),
Phys. Rev. {\bf D88}, 052002 (2013).

\end{thebibliography}
\end{document}